%% file: main_qpl.tex
\documentclass[submission,copyright,creativecommons,sharealike,noncommercial]{eptcs}
\pdfoutput=1

\input{preamble.tex}

\providecommand{\event}{QPL 2021}

\begin{document}

\title{CPM Categories for Galois Extensions}
\author{James Hefford
\institute{University of Oxford, Oxford, UK}
\email{james.hefford@cs.ox.ac.uk}
\and
Stefano Gogioso
\institute{Hashberg Quantum, London, UK}
\email{quantum(at)hashberg(dot)io}
}

\def\titlerunning{CPM Categories for Galois Extensions}
\def\authorrunning{J. Hefford \& S. Gogioso}

\maketitle

\begin{abstract}
  By considering a generalisation of the CPM construction, we develop an infinite hierarchy of probabilistic theories, exhibiting compositional decoherence structures which generalise the traditional quantum-to-classical transition.
  Analogously to the quantum-to-classical case, these decoherences reduce the degrees of freedom in physical systems, while at the same time restricting the fields over which the systems are defined.
  These theories possess fully fledged operational semantics, allowing both categorical and GPT-style approaches to their study.
\end{abstract}

\section{Introduction}

The CPM construction inhabits a prominent position in the study of Categorical Quantum Mechanics (CQM) \cite{abramsky_coecke}.
It is the natural categorical generalisation of the transition from pure quantum theory to mixed state quantum theory, taking a $\dag$-compact category $\cat{C}$ and producing a category $\cpm{\cat{C}}{}$ of completely positive maps
\cite{selinger_cpm,selinger_idempotents}.
These ``doubled'' categories are ubiquitous in the CQM community, for they provide the ideal setting in which to study finite-dimensional quantum theory: physically irrelevant global phases are cancelled out and the category allows for a natural description of the interface between quantum and classical theory.
Indeed, by considering the subcategory of the Karoubi envelope of $\cpm{\cat{C}}{}$ spanned by decoherence maps one can produce a category of C*-algebras, known elsewhere as the CP* construction, which unifies quantum theory with classical theory \cite{coecke_cp,Cunningham_cp,coecke_classicality}.

It was recently pointed out by one of the authors that the original CPM construction is a special case of a generalised, ``higher-order'' CPM construction \cite{gogioso_cpm}: from now on, when talking about CPM construction we shall refer to the latter, generalised version.
This generalisation expands upon the inherent symmetries of the original CPM construction to produce categories with more exotic properties, captured by an essential invariance under the action of a group of monoidal autofunctors.
These categories have found uses in quantum natural language processing, where they capture multiple degrees of linguistic ambiguity \cite{ashoush,piedeleu,coecke_double,coecke_dilation}, and in CQM, where they have been found to exhibit higher-order interference, hyper-decoherence and a rich phase group structure \cite{gogioso_hyper,hefford_hypercubes}.

In this work we show that these CPM categories exhibit rich decoherence structures, generalising that of the quantum-to-classical transition.
The decoherence maps compose as one might intuitively expect, giving towers of subcategories corresponding to intermediate CPM constructions.
We then focus on the particular case of CPM categories induced by finite degree field extensions and uncover some interesting connections between group theory and decoherence, with Galois theory as the bridge between them.
In doing so, we explicitly construct an infinite family of related probabilistic theories with interesting and complex decoherence towers.
This effort fits squarely within a recent surge of interest in hyper-decoherence \cite{zyczkowski_quartic,lee_interference,lee_nogo,gogioso_hyper,hefford_hypercubes} and resource theories of coherence \cite{Selby_coherence}.

The decoherences studied here are similar to the quantum-to-classical decoherence on two fronts:
\begin{itemize}
  \item They reduce the degrees of freedom of systems, analogous to sending a density matrix to a classical probability distribution
  \item They reduce the field over which the systems are defined, analogous to the transition from $\complex$ to $\reals$ \footnote{the positivity of the image of usual quantum-to-classical decoherence arises because $|z|^2 \geq 0$. This is generalised in this work by the field norm which we will show acts to restrict the image of decoherences to a sub-semifield of generalised positive elements}
\end{itemize}
It is this latter point where Galois theory will play its role; we will be able to develop towers of decoherences that mimic the lattice of sub-fields of a Galois extension.
In order to be able to study Galois extensions with non-Dedekind Galois groups we develop a further slight generalisation of the CPM construction of \cite{gogioso_cpm}, built from group transversals instead of orbits of entire groups.
We aim to closely investigate this further generalisation while connecting it to the literature on equivariant categories---that is, subcategories of morphisms fixed by a group action.
In particular, we show that the new CPM categories are hierarchically compositional: they can be formed step-wise by going up any tower of subgroups of the overall group action.

\section{CPM Categories}\label{sec:cpmcats}

In this section, we give an overview of the CPM construction from \cite{gogioso_cpm}.
We provide a further generalisation to foldings that are given by transversals of subgroups, and we connect the construction to work on equivariant categories.

The CPM construction from \cite{gogioso_cpm} can be understood as a two-step process.
Starting with a symmetric monoidal category $\cat{C}$ one:
\begin{enumerate}
  \item first, ``folds'' the objects and morphisms according to a strict left action by monoidal autofunctors $\phi:G \morph{}\aut{\cat{C}}$ for a finite group $G$:
  \begin{equation*}
      A \mapsto \bigotimes_{g \in G} \phi_{g} A
  \end{equation*}
  \item then, discards environment systems $E$ by applying effects $\xi_E$ from a collection $\Xi$ of monoidally closed ``discarding maps'':
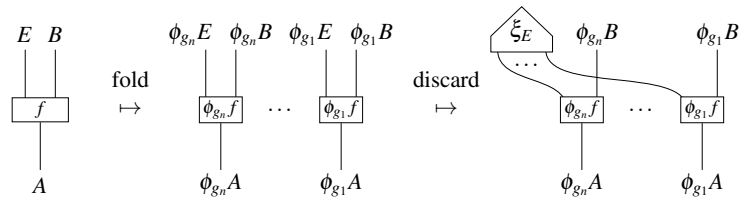
    \begin{equation*}
      \tikzfigscale{0.8}{figs/fold_discard}
    \end{equation*}
\end{enumerate}
This two-step process generalises the doubling and discarding from the original CPM construction \cite{selinger_cpm,selinger_idempotents}, where the discarding maps are chosen to be the counits of the compact closed structure.
The following two sections dive into the details of each step, starting with folding.

\subsection{Folding and Equivariant Categories}

Consider a symmetric monoidal category (SMC) $\cat{C}$ equipped with a strict left action by monoidal autofunctors $\phi:G\morph{}\aut{\cat{C}}$, for some finite group $G$.
We restrict our attention to strict actions, since it is known that any category with a weak $G$-action (i.e. one with isomorphisms $\phi_g\phi_h\simeq\phi_{gh}$ satisfying certain compatibility conditions) can be strictified and is equivalent to a category with a strict $G$-action $\phi_g\phi_h=\phi_{gh}$ \cite{shinder_groupactions}.
From such a SMC with a $G$-action, one can derive a category of $G$-equivariant morphisms \cite{elagin_equivariant, ganter_symmetric}, as follows.

\begin{defn}[$G$-equivariant Category]
  Let $\cat{C}$ be a category, $G$ be a finite group and $\phi:G\morph{}\aut{\cat{C}}$ be a strict left action.
  The objects of the \emph{$G$-equivariant category} $\cat{C}_G$ are pairs $(A,(\eta_A^g)_{g \in G})$ of an object $A$ of $\cat{C}$ and a family of isomorphisms $\eta_A^g: A\morph{}\phi_g A$ for each $g\in G$, such that the following diagram commutes for all $g,h\in G$:
  \begin{equation}\label{diag:eta}
    \begin{tikzcd}
      A \ar[r,"\eta_A^h"] \ar[rd,"\eta^{hg}_A"'] & \phi_h A \ar[d,"\phi_h\eta^g_A"]\\
      & \phi_h\phi_gA
    \end{tikzcd}
  \end{equation}
  The morphisms $f:(A,(\eta_A^g)_{g \in G})\morph{}(B,(\eta_B^g)_{g \in G})$ of $\cat{C}_G$ are the morphisms of $\cat{C}$ which commute with the isomorphisms $\eta^g_A$ for all $g\in G$:
  \begin{equation}\label{diag:equivariantmorphs}
    \begin{tikzcd}
      A \ar[d,"f"'] \ar[r,"\eta_A^g"] & \phi_g A  \ar[d,"\phi_g f"] \\
      B \ar[r,"\eta_B^g"'] & \phi_g B
    \end{tikzcd}
  \end{equation}
  There is a canonical forgetful functor $\iota:\cat{C}_G\morph{}\cat{C}$ forgetting the equivariant structure.
\end{defn}

\begin{prop}\label{prop:equivariantmonoidal}
  Let $\cat{C}$ be a SMC and denote by $\otimes$ its tensor product.
  Let $\cat{C}$ be equipped with a $G$-action by strict monoidal autofunctors.
  Then $\cat{C}_G$ is also symmetric monoidal, with tensor product $\boxtimes$ defined as follows:
  \begin{equation}
      (A,(\eta_A^g)_{g \in G})\boxtimes(B,(\eta_B^g)_{g \in G}) := (A\otimes B, (\eta_A^g\otimes\eta_B^g)_{g \in G})
  \end{equation}
\end{prop}
\begin{proof}
  Proof given in appendix \ref{proof:equivariantmonoidal}.
\end{proof}

For any subgroup $H\leq G$, the $G$-action $\phi$ descends by restriction to an $H$-action, so one can form the $H$-equivariant category $\cat{C}_H$.
It will often be the case that we have some preferred collection of the isomorphisms $\eta_A^h:A\morph{}\phi_h A$, chosen to be \emph{compatible with the monoidal product $\otimes$ of $\cat{C}$}, so that $\eta_A^h\otimes \eta_B^h = \eta_{A\otimes B}^h$ for all objects $A$ and $B$ and group elements $h\in H$.
We write $\hat{\cat{C}}_{H,\eta}$ for the full subcategory of $\cat{C}_H$ spanned by objects involving isomorphisms in this family, those of the form $(A,(\eta_A^h)_{h \in H})$.

\begin{prop}\label{prop:naturaliser}
  The category $\hat{\cat{C}}_{H,\eta}$ is equivalent to the naturaliser $\textrm{Nat}((\eta^h)_{h \in H})$ of the infranatural isomorphisms $\eta^h$ in $\cat{C}$, i.e. the largest subcategory of $\cat{C}$ such that all of the $\eta^h$ are natural (so that each autofunctor $\phi_h$ is naturally isomorphic to the identity).
\end{prop}
\begin{proof}
  Proof given in appendix \ref{proof:naturaliser}.
\end{proof}


\begin{defn}[Folding Functor]
  Let $G$ be any group and let $T$ be a left transversal of some subgroup $H \leq G$.
  Let $(\eta_a^h)_{h \in H}$ be a collection of isomorphisms compatible with the monoidal product.
  We refer to the tuple $\tau:=(G, H, \eta, T)$ as the \emph{folding data}, and we define the \emph{folding functor} $\fld{\tau}:\hat{\cat{C}}_{H,\eta}\morph{}\cat{C}$ as follows:
  \begin{equation}\label{folding}
    (A,(\eta_A^h)_{h \in H}) \mapsto \bigotimes_{t\in T} \phi_t \iota A
    \hspace{20mm}
    f \mapsto \bigotimes_{t\in T} \phi_t \iota f
  \end{equation}
\end{defn}

\begin{remark}
  The folding functor from \cite{gogioso_cpm} arises as the special case where $H$ is taken to be the trivial group.
  A transversal $T$ of $H:=\{e\}$ in $G$ is then precisely the set $T=G$ of all the elements of $G$.
  Since $\phi$ is strict we have $\phi_e=\id{\cat{C}}$ and hence $\cat{C}_{\{e\}} \simeq \cat{C}$, so that the folding functor can be considered an endofunctor $\fld{G}$ on $\cat{C}$.
  We call this a \emph{complete folding functor}, since it uses all the elements of its defining group.
\end{remark}

\begin{defn}[Folded Category]
  For given folding data $\tau:=(G, H, \eta, T)$, the \emph{folded category} $\FLD{\tau}(\cat{C})$ is the subcategory of $\cat{C}$ formed by the image of the folding functor, with objects and morphisms of the form \eqref{folding}.
  For a complete folding functor $\fld{G}$, we write $\FLD{G}(\cat{C})$.
\end{defn}

The folded category has pleasing symmetry properties - it is essentially invariant under each of the autofunctors $\phi_g$.
This follows because for any $g\in G$ the autofunctor $\phi_g$ acts on the indices of the tensor product to send a transversal $T$ to another transversal $T'$.
In general, this new transversal differs from the original in two ways - the ordering of the cosets to which each element belongs is permuted and the representative of each coset has been altered.
To recover the original transversal we can compose two isomorphisms.
Firstly, $\sigma^g$ which rearranges the cosets back to the original ordering, by a composition of the symmetry isomorphisms arising from the symmetric monoidal structure.
Secondly, $\rho^g$ which recovers the representatives of the original transversal and is given by a composition of the isomorphisms $\eta^g$.
Formally, $\rho_A^g$ is given by
\begin{equation*}
  \rho_A^g := \bigotimes\limits_{t\in T} \phi_t (\eta^{h_t}_A)^{-1}
\end{equation*}
where each $h_t$ is the element of $H$ which separates the representatives of a given coset of $H$ between the two transversals $T$ and $T'$ \footnote{That is, if $t\in T$ is the representative of a coset $tH$ then in the transversal $T'$, the representative of $tH$ is given by $th_t$ for some $h_t\in H$.}
The following proposition establishes this formally:

\begin{prop}\label{prop:autofunctors_folded}
  The autofunctors $\phi_g: \cat{C} \rightarrow \cat{C}$ restrict to functors $\phi_g: \FLD{\tau}(\cat{C}) \rightarrow \FLD{\tau}(\cat{C})$ on the folded category, and are naturally isomorphic to the identity on $\FLD{\tau}(\cat{C})$.
\end{prop}
\begin{proof}
  Proof given in appendix \ref{proof:autofunctors_folded}.
\end{proof}

\begin{remark}
  For a complete folding functor, the previous proposition shows that there is a canonical embedding of $\FLD{G}(\cat{C})$ into a $G$-equivariant category where the structural isomorphisms $\eta^g$ are given by the permutations $\sigma^g$ arising from compositions of the symmetry isomorphisms from the symmetric monoidal structure.
  This remark will be useful for the following propositions.
\end{remark}

\begin{prop}\label{prop:foldfactor}
  Suppose we have a complete folding functor for the $G$-action $\phi$ and suppose there is a subgroup $H\leq G$.
  Then the complete folding functor factorises into a complete folding functor for the induced $H$-action followed by the folding functor for any choice of transversal $T$ of $H$ in $G$.
\end{prop}
\begin{proof}
  Proof given in appendix \ref{proof:foldfactor}.
\end{proof}

\noindent
When $H\trianglelefteq G$ is normal in $G$, we might expect that a complete folding functor for the $G$-action factorises through a \emph{complete} folding functor for the quotient $G/H$-action.
The following proposition establishes that this is indeed the case.

\begin{prop}\label{prop:foldfactor2}
  Suppose we have a complete folding functor for the $G$-action $\phi$.
  Let $H\trianglelefteq G$ be a normal subgroup of $G$.
  Then the complete folding functor for the $G$-action factorises through complete folding functors for the $H$-action and the quotient $G/H$-action.
\end{prop}
\begin{proof}
  Proof given in appendix \ref{proof:foldfactor2}.
\end{proof}

\subsection{Environment Structures}

The folded category $\FLD{\tau}(\cat{C})$ forms the starting point for a family of CPM categories.
The second ingredient is the choice of environment structure.

\begin{defn}[Environment Structure]
  An environment structure $\Xi$ is a family of sets $\Xi_A$ of effects $\xi_A:\fld{\tau}A\morph{}I$ for each object $A$ of $\cat{C}$ satisfying the following conditions:
  \begin{itemize}
    \item $\xi_A\in\Xi_A, \xi_B\in \Xi_B \implies (\xi_A\otimes \xi_B) \circ \pi_{A,B}^{-1} \in \Xi_{A\otimes B}$
    \item $\Xi_I = \{\id{I}\}$
    \item $\xi_A\in \Xi_A, g\in G \implies \phi_g\xi_A = \xi_A \circ {\left(\sigma_A^g\right)}^{-1} \circ {\left(\rho_A^g\right)}^{-1}$
  \end{itemize}
  where $\sigma_A^g$ and $\rho_A^g$ are the isomorphisms defined in the previous section and $\pi_{A,B}$ is the following permutation (obtained by composition of symmetry isomorphisms):
  \begin{equation*}
    \pi_{A,B}:\bigotimes_{t\in T}\phi_t A \otimes \bigotimes_{t\in T} \phi_t B \longrightarrow \bigotimes_{t\in T} \phi_t (A\otimes B)
  \end{equation*}
\end{defn}

\noindent
The constraints of the environment structure ensure that $\Xi$ is essentially closed under the monoidal product and that all the effects are invariant under the action of the autofunctors $\phi_g$, up to (well-behaved) natural isomorphism.

\begin{remark}
  What has previously been called an environment structure in the literature  \cite{coecke_environment,coecke2016terminality} coincides with the previous definition when each set $\Xi_A$ contains exactly one element.
  Indeed, it is necessary to pick a particular effect for each object $A$ (with the choices being closed under the monoidal product) to act as the designated overall discarding maps $\{\Ground_A\}$, bestowing the theory with a notion of causality.
\end{remark}

\begin{defn}[CPM Category]
  The \emph{CPM category} $\cpm{\cat{C}}{\tau,\Xi}$ is the smallest subcategory of $\cat{C}$ containing the folded category $\FLD{\tau}(\cat{C})$, all the effects of $\Xi$ and their monoidal products with the identity morphisms, defined as follows:
  \begin{equation*}
    \xi_A\boxtimes\id{\fld{\tau}B}
    :=
    (\xi_A\otimes\id{\fld{\tau}B}) \circ \pi_{A,B}^{-1}
  \end{equation*}
\end{defn}

\begin{remark}
  The CPM category is essentially invariant under the autofunctors, because so are both the folded category and the effects in the environment structure.
\end{remark}

\begin{prop}\label{prop:cpmmonoidal}
  The CPM category $\cpm{\cat{C}}{\tau,\Xi}$ is a symmetric monoidal category when equipped with the following tensor product, having $\FLD{\tau}(\cat{C})$ as a monoidal subcategory:
  \begin{equation}
    F \boxtimes G
    :=
    \pi_{C, D} \circ (F \otimes G) \circ \pi_{A,B}^{-1}
  \end{equation}
  where $F: \fld{\tau}A \rightarrow \fld{\tau}C$ and $G: \fld{\tau}B \rightarrow \fld{\tau}D$ are generic morphisms in $\cpm{\cat{C}}{\tau,\Xi}$.
\end{prop}
\begin{proof}
  Proof given in appendix \ref{proof:cpmmonoidal}.
\end{proof}

\begin{prop}\label{prop:cpmdagger}
  Let $\cat{C}$ be a $\dag$-compact category equipped with a monoidal $G$-action $\phi$.
  If there is some $g \in G$ such that $\phi_g=\conj{\cat{C}}$ is the conjugating autofunctor, then $\cpm{\cat{C}}{\tau,\Xi}$ is a $\dag$-compact category.
\end{prop}
\begin{proof}
  Proof given in appendix \ref{proof:cpmdagger}.
\end{proof}

We conclude this section by recalling a few simple examples of CPM categories that have already appeared in the literature: the original CPM construction, density hypercubes, double dilation and double mixing.

\begin{example}
  Taking $\cat{C}$ to be $\dag$-compact, the group action $\phi:C_2\morph{}\aut{\cat{C}}$ to send $ 0\mapsto \id{\cat{C}}, 1\mapsto \conj{\cat{C}}$ and the transversal to be the whole group $T=C_2$, we obtain the folding (aka doubling) for the original CPM construction.
  The original CPM construction is then recovered by taking $\Xi_A$ to contain only the caps for each object $A$ of $\cat{C}$.
\end{example}

\begin{example}
  By taking $\phi:C_2\times C_2 \morph{}\aut{\cat{C}}$ to send $(0,1)\mapsto \conj{\cat{C}}, (1,0)\mapsto \conj{\cat{C}}$, with $T=C_2\times C_2$, we recover the folding of density hypercubes \cite{gogioso_hyper,hefford_hypercubes}, double dilation \cite{coecke_dilation} and double mixing \cite{ashoush,coecke_double,piedeleu}.
  In all three cases the environment structure contains the caps from the original CPM construction (i.e. its discarding maps).
  For density hypercubes one additionally picks the caps induced by a special commutative $\dag$-Frobenius algebra; for double dilation the caps induced by the action of the folding functor on the caps of the original CPM construction; and for double mixing a spider induced by a special commutative $\dag$-Frobenius algebra.
  Note that the additional caps for density hypercubes and double-dilation are wired differently, see discussion in \cite{hefford_hypercubes}.
\end{example}

\subsection{Environment Structures from Classical Structures}

Suppose now that we are working in a symmetric monoidal $\dag$-category $\cat{C}$ which is rich in special commutative $\dag$-Frobenius algebras ($\dag$-SCFAs, also known as \emph{classical structures}). In particular, assume that each object $A$ has at least one $\dag$-SCFA on it.
Classical structures provide an analogy to the category $\fhilb$ of finite dimensional Hilbert spaces, where $\dag$-SCFAs are in bijection with bases of a given Hilbert space \cite{coecke_bases}.
These $\dag$-SCFAs are of vital importance to the categorical study of quantum theory, where they capture, to name but a few, phases \cite{coecke_zx}, Fourier transforms \cite{gogioso2015fourier} and decoherence.
It is the latter that we will focus on as part of this work.

In the original CPM construction, decoherence can be studied by forming the Karoubi envelope and then taking a full subcategory spanned by so-called \emph{decoherence maps}.
For any $\dag$-SCFA, one can use the $\dag$-compact closure of $\fhilb$ to form the following map:
\begin{equation}\label{eq:conjspider}
  \tikzfigscale{0.8}{figs/conjspider}
\end{equation}
The decoherence maps are given by the composition $\dec{}{}:=\delta^\dag\circ\delta$.
One can show that the full subcategory of the Karoubi envelop $\splt{\cpm{\cat{\fhilb}}{}}$ spanned by objects of the form $(H^*\otimes H,\dec{}{})$ is equivalent to $\mat{\reals^+}$, the category of positive real valued matrices, while of course the full subcategory spanned by objects of the form $(H^*\otimes H,\id{})$ is equivalent to $\cpm{\fhilb}{}$.
Consequently, the full subcategory of $\splt{\cpm{\fhilb}{}}$ spanned by both decoherence maps and identities captures just enough to have both quantum theory and classical theory live within the same categorical setting.
The morphism $\delta$ from \eqref{eq:conjspider} can be used to construct an isomorphism $\theta:H^*\morph{}H$, allowing $\delta$ to be re-written as:
\begin{equation*}
  \tikzfigscale{0.8}{figs/spideriso} \hspace{3cm}
  \tikzfigscale{0.8}{figs/conjspider2}
\end{equation*}
This is to be expected, since the existence of a $\dag$-SCFA on an object $H$ implies that $H$ is self-dual.
Looking at this from the perspective of higher-order CPM constructions, this suggests that it could be interesting to consider categories $\cat{C}$ equipped with a choice of isomorphisms $\theta_A^g:A\morph{}\phi_g A$ for each object $A$ and element $g\in G$, subject to suitable conditions; together with these isomorphisms, the presence of $\dag$-SCFAs implies that $A$ is dual to $\phi_g A$ for all $g\in G$.

\begin{defn}[Generalised Classical Structures]
  Let $\cat{C}$ be a $\dag$-compact category equipped with a $G$-action $\phi$.
  A \emph{generalised classical structure} on an object $A$ of $\cat{C}$ is a pair $(\hbox{\input{symbols/ZbwdotSym.tex}}\!\!, \theta_A)$ of a $\dag$-SCFA $\hbox{\input{symbols/ZbwdotSym.tex}}\!\!$ on $A$ and a choice of automorphisms $\theta = (\theta^g)_{g \in G}$ on $A$ satisfying the same commutative diagram \eqref{diag:eta} as $\eta_A^{g}$ together with the following compatibility conditions:
  \begin{equation*}
    \phi_g \big( \tikzfigscale{0.8}{figs/gen_class_struct1} \big) \hspace{5mm} = \hspace{5mm} \tikzfigscale{0.8}{figs/gen_class_struct2}
    \hspace{20mm}
    \phi_g \big( \tikzfigscale{0.8}{figs/gen_class_struct3} \big) \hspace{5mm} = \hspace{5mm} \tikzfigscale{0.8}{figs/gen_class_struct4}
  \end{equation*}
\end{defn}

\begin{defn}[Complete Decoherence Maps]
  \label{def:complete-decoh-discard-maps}
  Let $\cat{C}$ be a $\dag$-compact category equipped with a $G$-action $\phi$.
  The \emph{complete decoherence map} for a generalised classical structure $(\hbox{\input{symbols/ZbwdotSym.tex}}\!\!, \theta)$ is the morphism of $\mathcal{C}$ defined as follows:
  \begin{equation}\label{eq:complete-decoh}
    \tikzfigscale{0.8}{figs/decoh}
    \hspace{5mm}:=
    \tikzfigscale{0.8}{figs/genspider}
  \end{equation}
  This is also a morphism in any CPM category $\cpm{\cat{C}}{G,\Xi}$ where the environment structure $\Xi$ contains the following \emph{complete discarding map} $(\hbox{\input{symbols/ZbwdotSym.tex}}\!\!, \theta)$:
  \begin{equation}\label{eq:complete-discard}
    \tikzfigscale{0.8}{figs/groupdiscard}
  \end{equation}
  In this case, the spider part of the leftmost diagram in \eqref{eq:complete-decoh} is simply the $G$-folding $\fld{G}(\!\hbox{\input{symbols/ZbwcomultSym.tex}}\!\!)$ of the comultiplication for the $\dagger$-SCFA $\hbox{\input{symbols/ZbwdotSym.tex}}\!\!$.
\end{defn}

In \eqref{eq:complete-decoh} and \eqref{eq:complete-discard}, the diagrams with thick lines are in the graphical calculus for the $\dag$-compact category $\cpm{\cat{C}}{G,\Xi}$, while the diagrams with thin lines are in the graphical calculus for the $\dag$-compact category $\cat{C}$.
This follows the same convention as diagrams for the original CPM construction.
Unlike the discarding maps from the original CPM construction, however, the complete discarding maps defined by \eqref{eq:complete-discard} depend on a choice of generalised classical structure $(\hbox{\input{symbols/ZbwdotSym.tex}}\!\!, \theta)$.
This dependence is left implicit in our diagrammatic notation.

Recall, now, the result from Proposition \ref{prop:foldfactor}: if $H \leq G$, then the complete $G$-folding factors into the complete $H$-folding followed by the folding over any transversal $T$ of $H$ in $G$.
Via this mechanism, we can define discarding maps and decoherence maps corresponding to all possible choices of $H \leq G$ and transversal $T$.

\begin{defn}[$H$-Decoherence Maps]
  Let $\cat{C}$ be a $\dag$-compact category equipped with a $G$-action $\phi$.
  Let $H \leq G$ be a subgroup, $T$ be a transversal for $H$ in $G$, and $\tau := (G, H, \eta, T)$ be folding data.
  The \emph{$H$-decoherence map} for a generalised classical structure $(\hbox{\input{symbols/ZbwdotSym.tex}}\!\!, \theta)$ and folding data $\tau$ is the morphism of $\mathcal{C}$ defined as follows:
  \begin{equation}\label{eq:decoh-subgroup}
    \tikzfigscale{0.8}{figs/tau-decoh}
    \hspace{5mm}:=\hspace{5mm}
    \fld{\tau}\Big(
    \tikzfigscale{0.8}{figs/h-decoh}
    \Big)
  \end{equation}
  This is also a morphism in any CPM category $\cpm{\cat{C}}{G,\Xi}$ where the environment structure $\Xi$ contains the following \emph{$H$-discarding map} $\Ground_{\hspace{-1mm}H}: \fld{G} A \rightarrow I$ for $(\hbox{\input{symbols/ZbwdotSym.tex}}\!\!, \theta)$ and $\tau$:
  \begin{equation}\label{eq:discard-subgroup}
    \tikzfigscale{0.8}{figs/tau-groupdiscard}
    \hspace{5mm}:=\hspace{5mm}
    \fld{\tau}\Big(
    \tikzfigscale{0.8}{figs/h-groupdiscard}
    \Big)
  \end{equation}
\end{defn}

The dependence of $H$-decoherence maps and $H$-discarding maps on both the generalised classical structure $(\hbox{\input{symbols/ZbwdotSym.tex}}\!\!, \theta)$ and the folding data $\tau$ is left implicit in the diagrammatic notation: the subgroup $H$ is the only piece of data that will be of practical interest.
Furthermore, by choosing $H := G$ we always recover the complete decoherence maps and discarding maps from Definition \ref{def:complete-decoh-discard-maps}.

It is a straightforward observation, following from the Spider Theorem \cite{coecke_kissinger_2017, heunen_categories, lack_props}, that the $H$-decoherence maps are idempotent and causal with respect to the corresponding $H$-discarding maps:
\begin{equation}
  \tikzfigscale{0.8}{figs/dec_idemp}
\end{equation}
A similarly straightforward observation is that decoherence maps for subgroups are compositionally well-behaved: if $\dec{}{H}$ and $\dec{}{K}$ are the $H$-decoherence and $K$-decoherence maps for two subgroups $H,K\leq G$, then the following is true, where $H\vee K$ is the group theoretic join of the two subgroups:
\begin{equation}
  \dec{}{H}\circ\dec{}{K} = \dec{}{K}\circ\dec{}{H} = \dec{}{H\vee K}
\end{equation}

Now, suppose that we fix some subgroup $H \leq G$, folding data $\tau = (G, H, \eta, T)$ and that we equip each object of $\cat{C}$ with a choice of generalised classical structures $\Gamma := \left((\hbox{\input{symbols/ZbwdotSym.tex}}\!\!_A, \theta_A)\right)_{A \in \operatorname{obj}\,\cat{C}}$ on all objects, compatibly with its monoidal structure.

\begin{prop}\label{prop:H-environment}
  Under the conditions above, we obtain an environment structure $\Xi$ by associating to each object $A \in \operatorname{obj}\,\cat{C}$ the singleton set $\Xi_A$ containing only the $H$-discarding map.
\end{prop}
\begin{proof}
  Proof given in appendix \ref{proof:H-environment}.
\end{proof}

The environment structure obtained above can be seen to generalise the standard choice in the original CPM construction: in that case, the group $G=C_2$ only allows a single non-trivial choice, the one corresponding to $H=C_2$ itself.
As for double dilation, double mixing and density hypercubes, the group is $G = C_2\times C_2$ and the $G$-action is defined by:
\begin{equation}
  \phi_{(a,b)} :=
  \begin{cases}
    \operatorname{id}_{\cat{C}} & \text{ if } a \oplus b = 0\\
    \conj{\cat{C}} & \text{ if } a \oplus b = 1
  \end{cases}
\end{equation}
All three models contain the $H$-discarding maps for the subgroup $H := \{(0, 0), (0, 1)\} \cong C_2$, while the other effects are $K$-discarding maps for different choices of subgroup $K$:
\begin{align*}
  \text{density hypercubes}, \ K_{dh} := \{(0, 0), (1, 1)\}; \qquad & \text{double dilation}, \ K_{dd} := \{(0, 0), (1, 0)\}; \\
  \text{double mixing}, & \ K_{dm} := G;
\end{align*}
The previously discussed behaviour of discarding maps under composition then generalises the result of Zwart and Coecke \cite{coecke_dilation} showing that double mixing is a sub-theory of double dilation, and the result of Gogioso and Scandolo \cite{gogioso_hyper} showing that double mixing is a sub-theory of density hypercubes:
\begin{equation*}
  K_{dd} \vee H = G = K_{dm}
  \hspace{5mm}\Rightarrow\hspace{5mm}
  \dec{}{H}\circ\dec{}{K_{dd}} = \dec{}{K_{dd}}\circ\dec{}{H} = \dec{}{K_{dm}}
\end{equation*}
\begin{equation*}
  K_{dh} \vee H = G = K_{dm}
  \hspace{5mm}\Rightarrow\hspace{5mm}
  \dec{}{H}\circ\dec{}{K_{dh}} = \dec{}{K_{dh}}\circ\dec{}{H} = \dec{}{K_{dm}}
\end{equation*}
A limitation of this group theoretic approach is that it does not tell us whether double dilation and density hypercubes coincide; by other means, it is in fact known that they do not \cite{hefford_hypercubes}.
This also means that theories generated by the same folding and using discarding maps from isomorphic subgroups of $G$ do not, in general, need to be equivalent: the $G$-action itself plays an important role in this.
An investigation of the exact conditions under which such theories are equivalent is left to future work.

\section{CPM Categories Induced by Galois Extensions}

We now get to the core of this work: building CPM categories from field extensions of finite degree.
A previous remark by one of the authors \cite{gogioso_cpm}, reported below, highlights why we take this route.

\begin{remark}
  Consider a SMC $\cat{C}$ which is enriched in commutative monoids, i.e. one where homsets have a commutative monoid structure $(+, 0)$ compatible with composition and tensor product.
  The scalars for such a category always form a semiring.
  Now assume that the scalars form a field $K$ and that the category is equipped with a $G$-action $\phi$ by linear monoidal autofunctors, i.e. autofunctors respecting the enrichment.
  Then the $\phi_g$ restrict to an action of $G$ on $K$ by field homomorphisms, and we can consider the fixed field $k$ for the action: because if $k$ is the fixed field for a finite subgroup of the automorphisms of $K$, the extension $k \subset K$ is Galois.
  The action of the complete folding functor $\fld{G}$ on scalars of $\cat{C}$ then coincides with a power of the field norm:
  \begin{equation*}
    \fld{G}(x) = \prod_{g \in G} \phi_g(x)
    =
    \left(
      \prod_{\xi \in \gal{K}{k}}\xi(x)
    \right)^{\frac{|G|}{|\gal{K}{k}|}}
  \end{equation*}
  If the $G$-action is faithful, i.e. if $G$ is bijected with $\gal{K}{k}$ by the action, then the exponent is 1 and the folding functor coincides with the field norm.
\end{remark}

In this section, we develop the connection between Galois theory and CPM constructions more fully.
Let $\mat{S}$ denote the category of matrices over a commutative semiring $S$, with the positive natural numbers $n \in \mathbb{N}^+$ as objects and the $n\times m$ matrices with entries from $S$ as morphisms $m\morph{}n$.
It is well known that $\mat{S}$ is a symmetric monoidal category enriched in commutative monoids, with a wealth of additional structure (e.g. biproducts).
If $k\subset K$ is a Galois extension, then its Galois group $\Gamma := \gal{K}{k}$ induces the following action on $\mat{K}$ by linear monoidal autofunctors:
\begin{itemize}
  \item on objects, $\phi_\gamma(A) := A$ for all $\gamma \in \Gamma$ and all objects $A$;
  \item on morphisms, $\phi_\gamma \left( (M_{i,j})_{j=1,...,m}^{i=1,...,n} \right) := \left(\gamma(M_{i,j})\right)_{j=1,...,m}^{i=1,...,n}$
\end{itemize}
As previously remarked, the complete folding functor $\fld{\Gamma}$ for this action acts as the field norm on the scalars of the category:
\begin{equation}
  \fld{\Gamma}(x) = N_{K/k}(x)
\end{equation}

\begin{defn}[Galois CPM Category]
  Let $k \subset K$ be a Galois extension with Galois group $\Gamma$.
  The \emph{Galois CPM category} associated to this extension is the CPM category $\cpm{\mat{K}}{K/k}:=\cpm{\mat{K}}{\Gamma,\Xi}$ obtained by considering the $\Gamma$-action induced by the Galois group on $\mat{K}$ and the complete folding functor $\fld{\Gamma}$.
  The environment structure $\Xi$ is obtained by taking the $H$-discarding maps for all subgroups $H \leq \Gamma$:
  \begin{equation*}
    \Xi_n := \big\{ \Ground_{\hspace{-1mm}H}: \fld{\Gamma}n \rightarrow I \,\big|\, H \leq \Gamma \big\}
  \end{equation*}
  with respect to the classical structure $\hbox{\input{symbols/ZbwdotSym.tex}}\!\!$ defined by the standard orthonormal basis $\left(|i\rangle\right)_{i=1,...,n}$ and closing the sets $\Xi_n$ under the tensor product.
\end{defn}

For any sub-semiring $S\subset K$ we also introduce the notation $\cpm{\mat{S}}{K/k}:=\cpm{\mat{S}}{\Gamma,\Xi}$ for the restriction of the Galois CPM category $\cpm{\mat{K}}{K/k}$ to the semiring $S$.
The category has the same folding and environment structure as the Galois CPM category but on $\mat{S}$ embedded into $\mat{K}$.
Note that because the folding is complete, and thus contains all possible embeddings of $K$ into its algebraic closure, the choice of embedding of $S$ into $K$ is not important.

Now, for any Galois CPM category, because the folding functor acts as the field norm, the pure scalars $\fld{\Gamma}(x)$ are always elements of the base field $k$.
Moreover, because $k$ is closed under addition, the mixed scalars which are obtained by applying complete discarding maps, i.e. those in the form:
\begin{equation*}
  \Ground \circ \fld{\Gamma}\left(\ket{v}\right)
  =
  \sum_{i=1}^n
  \fld{\Gamma}(v_i)
\end{equation*}
are also elements of $k$.

Similarly, for generic scalars obtained by using discarding maps for intermediate subgroups $\{1\}< \Lambda < \Gamma$, we can write the following:
\begin{equation}\label{genericscalar}
  \Ground_{\hspace{-1mm}\Lambda} \circ \fld{\Gamma}\left(\ket{v}\right)
  =
  \prod_{t \in T}
  \left(
  \sum_{i=1}^n
  \prod_{\mu \in t\Lambda}
    \phi_\mu(v_i)
  \right)
  = \prod_{t\in T} \phi_t \left(\sum_i \prod_{\lambda\in\Lambda} \phi_\lambda(v_i) \right) = N_{F/k}\left( \sum_i N_{K/F}(v_i) \right)
\end{equation}
Since for any $v_i\in K$, $N_{K/F}(v_i) \in F$ and $F$ is closed under addition, we have $\alpha:=\sum_i N_{K/F}(v_i) \in F$.
Consequently $N_{F/k}(\alpha) \in k$.
Indeed, any generic scalar of the Galois CPM category must be an element of the base field because it is fixed by all elements of the $\Gamma$-action, just as all morphisms are.
As noted, the automorphisms of the action descend on the scalars to the Galois automorphisms of $K$ by $k$.

It is not necessarily the case that the set $\textrm{End}(I)$ of all scalars of the Galois CPM category is isomorphic to $k$ itself but rather it must form a sub-semiring.
We have the additive and multiplicative units, 0 and 1, since they are pure scalars and closure under multiplication follows immediately from the fact we are working in a monoidal category.
The complete discarding maps $\Ground$ act to allow us to take arbitrary sums of any chosen scalars (e.g. by appending ancillary systems) and consequently $\textrm{End}(I)$ is also closed under addition.

It can be difficult in general to decide precisely which sub-semiring of $k$ the scalars form.
For instance we see from equation \eqref{genericscalar} that we will have to consider not only the closure of the norm from the top field to the base field, but also the iterative closure of the norm for each intermediate field where we take the norm to an intermediate $F$, close it under addition, and then take the norm of any element of this to the base field.

Nevertheless, in section \ref{sec:examples} we will see that for many large classes of extensions we can say fair amount about the semiring of scalars and in many cases fully characterise it.

\subsection{Decoherence Structures}

By construction, Galois CPM categories come with a family of decoherence maps which mimic the decoherence of quantum theory to classical theory.
Each subgroup $\Lambda$ of the Galois group $\Gamma$ induces a decoherence map via its corresponding effect in the environment structure $\Xi$.
The similarity with quantum theory is two-fold:
\begin{enumerate}
  \item The decoherence maps reduce the $\Gamma$-folding to one given by a transversal of the subgroup in the overall group.
  In the case of a normal subgroup, this gives a full folding by the quotient $\Gamma/\Lambda$.
  This serves to kill-off ``interference terms'' as one moves down the tower of subgroups, progressively reducing the degrees of freedom that can be used to describe the state of a system.
  This is akin to how traditional decoherence kills-off non-diagonal terms in the transition from density matrices to classical probability distributions.
  \item The decoherence maps also reduce the degree of the field extensions, producing theories over a series of sub-fields given by the Galois correspondence.
  This is akin to the transition from $\complex$ to $\reals$ in quantum theory.
\end{enumerate}

The second point follows immediately by considering the symmetries of any decohered map.
For simplicity, consider a state $v$ of a Galois CPM category.
We can expand the $H$-decohered $v$ in the orthonormal basis $\left(\ket{i_n}\right)_{i_n}$ as

\begin{equation*}
  \tikzfigscale{0.8}{figs/hdec_symmetries}
\end{equation*}

One sees that a generic term of this state is invariant under all autofunctors $\phi_h$ for $h\in H$ while the remaining autofunctors do not, in general, fix it.
Thus the entries of the matrix for the decohered $v$ are fixed by all $h\in H$, i.e. a subgroup of the Galois group and thus must belong to the corresponding intermediate field of the extension.

The hard part, in exactly the same way as for the scalars, is characterising how much of the intermediate field a given decoherence hits, since this depends on properties of the fields in question.
Nevertheless, it is clear that it is always a sub-semiring of the intermediate field.
We will study many examples in the next sections where, for large classes of extensions, we will be able to fully characterise the sub-semirings forming the images of the decoherences.

\subsection{Examples}\label{sec:examples}

\subsubsection{Number Fields}\label{sec:numberfields}
An interesting class of Galois extensions to consider are those of \emph{number fields}, finite algebraic extensions of $\rationals$.
This section is largely a compilation of results, with proofs in the appendix, the aim being to constrain the semirings at each level of decoherence.
We write $\N{K}{k}$ for the field norm $K\morph{}k$, $\clN{K}{k}$ for the additive closure of image of this norm (i.e. a sub-semifield of $k$) and $K^+$ for the set of totally positive elements of $K$.

It is a consequence of Waring's problem for fields \cite{ellison1,ellison2}, theorem \ref{thm:waring} (see appendix \ref{app:numfields}), that for any Galois extension $k\subset K$, all totally positive elements of $k$ are contained in the closure of the image of the norm from $K$ to $k$.
\begin{prop}\label{prop:pos_in_norm}
  Let $k\subset K$ be a finite extension with Waring's problem holding in $k$.
  Then $k^+\subset \clN{K}{k}$.
\end{prop}
\begin{proof}
  Proof given in appendix \ref{proof:pos_in_norm}.
\end{proof}

In particular, $k^+\subset \clN{K}{k}$ for all number fields since Waring's problem always holds.
It is worth pointing out that the upper bound $g(K,d)$ on the number of terms needed in Waring's problem places an upper bound on the dimension of ancillary systems we require to form all the totally positive elements of $k$ in any Galois CPM category.

A simple implication of proposition \ref{prop:pos_in_norm} is that when $k$ is not formally real, the closure of the norm coincides with $k$.

\begin{cor}\label{cor:nonreal}
  Let $k\subset K$ be a finite extension where $k$ has characteristic 0 and is not formally real.
  Then $\overline{N}_{K/k} = k$.
\end{cor}
\begin{proof}
  Proof given in appendix \ref{proof:pos_in_norm}.
\end{proof}

\begin{remark}
  If $k \subset K$ is a Galois extension then corollary \ref{cor:nonreal} implies that, in $\cpm{\mat{K}}{K/k}$, a decoherence map to an intermediate non-real field $L$ is full, in the sense that there is an equivalence of categories between the full subcategory of $\splt{\cpm{\mat{K}}{K/k}}$ spanned by the corresponding decoherence maps and $\cpm{\mat{L}}{L/k}$.
\end{remark}

\begin{prop}\label{prop:imagtoreal}
  Let $\rationals\subset k \subset K$ be a tower of extensions where $K$ and $k$ are both Galois over $\rationals$.
  If $K$ is totally imaginary and $k$ is totally real then $\clN{K}{k}=k^+$.
\end{prop}
\begin{proof}
  Proof given in appendix \ref{proof:imagtoreal}.
\end{proof}

\begin{cor}\label{prop:totimag_norm}
  Let $\rationals\subset K$ be a totally imaginary Galois extension.
  Then $\overline{N}_{K/\rationals} = \rationals^+$
\end{cor}

\begin{remark}
Corollary \ref{prop:totimag_norm} shows that the scalars of the folded category $\FLD{\Gamma}(\mat{K})$ for totally imaginary $K$ are always elements of $\rationals^+$.
\end{remark}

We can also go some way to dealing with decoherences from totally real Galois number fields.

\begin{prop}\label{prop:totreal_norm}
  Let $\rationals\subset K$ be a totally real Galois extension.
  Then $\clN{K}{\rationals} = \rationals$.
\end{prop}
\begin{proof}
  Proof given in appendix \ref{proof:totreal_norm}.
\end{proof}

\begin{remark}
  Proposition \ref{prop:totreal_norm} shows that for CPM categories generated by complete foldings for totally real Galois number fields, the mixed scalars arising from the complete discarding map are enough to capture all of $\rationals$, and thus $\textrm{End}(I)\simeq \rationals$.
\end{remark}

Characterising general decoherences between totally real fields is left to future work.

\subsubsection{Cyclotomic Extensions}

Cyclotomic extensions are obtained by adjoining a primitive $n^\textrm{th}$ root of unity $\zeta_n$ to the rationals.
These extensions are always Galois, since they are the splitting fields of the cyclotomic polynomials $\Phi_n(x)$.
They are of interest to us because they give easily constructible examples of Galois extensions with abelian Galois groups, isomorphic to $(\integers/n\integers)^\times$.
Moreover, the Kronecker-Weber theorem guarantees that every finite abelian extension of the rational numbers is a subfield of some cyclotomic field.

All cyclotomic extensions are totally imaginary: they are given by an imaginary quadratic extension of the totally real field $\rationals(\zeta_n+\zeta_n^{-1})$.
This means that they inherit the results of section \ref{sec:numberfields}: in particular we can constrain decoherences with corollary \ref{cor:nonreal} and proposition \ref{prop:imagtoreal}.

Let us now present our first explicit example of a Galois CPM theory.
Consider the cyclotomic extension $\rationals\subset \rationals(\zeta_5)$ where $\zeta_5$ is a primitive fifth root of unity.
As discussed in examples \ref{ex:cyclo1} and \ref{ex:cyclo2}, the Galois group of this extension is $\Gamma:=\langle\sigma\mid\sigma^4=\id{}\rangle\simeq C_4$, where $\sigma::\zeta_5\mapsto\zeta_5^2$. The Galois group has a single non-trivial subgroup $\Lambda:=\langle\sigma^2\rangle\simeq C_2$.
This subgroup is in correspondence with the field $\textrm{Fix}(\Lambda) = \rationals(\sqrt{5})$, which is a Galois extension of $\rationals$ since $\Lambda$ is normal in $\Gamma$.

We have the following equalities of closures of norms consistent with the results of section \ref{sec:numberfields}:
\begin{equation*}
  \clN{\rationals(\zeta_5)}{\rationals} = \rationals^+ \hspace{4cm} \clN{\rationals(\zeta_5)}{\rationals(\sqrt{5})} = \rationals(\sqrt{5})^+
\end{equation*}

The states of $\cpm{\mat{\rationals(\zeta_5)}}{\rationals(\zeta_5)/\rationals}$ take the following form:
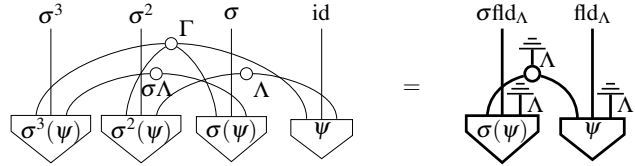
\begin{equation*}
  \tikzfigscale{0.8}{figs/fifthcyclo_state}
\end{equation*}
where the discarding maps on the left have been labelled by the cosets that induce them.
On the left we have used the string diagrams of $\mat{\rationals(\zeta_5)}$ and on the right those of $\cpm{\mat{\rationals(\sqrt{5})^+}}{\rationals(\sqrt{5})/\rationals}$.
There are two decoherence maps, corresponding to the subgroups $\Lambda$ and $\Gamma$ respectively:
\begin{equation*}
  \tikzfigscale{0.8}{figs/fifthcyclo_halfdec} \qquad \qquad \tikzfigscale{0.8}{figs/fifthcyclo_fulldec}
\end{equation*}

\begin{prop}\label{prop:cyclo5}
  Define $\quant_{\rationals(\zeta_5)/\rationals}:= \splt{\cpm{\mat{\rationals(\zeta_5)}}{\rationals(\zeta_5)/\rationals}}$ to be the Karoubi envelope for the Galois CPM category from the previous example.
  The full sub-categories of $\quant_{\rationals(\zeta_5)/\rationals}$ spanned by the decoherence maps are characterised as follows:
  \begin{itemize}
    \item $\dec{\rationals(\zeta_5)}{\rationals}$ decoherence maps
    $\longrightarrow$
    equivalent to $\mat{\rationals^+}$
    \item $\dec{\rationals(\zeta_5)}{\rationals(\sqrt{5})}$ decoherence maps
    $\longrightarrow$
    equivalent to $\cpm{\mat{\rationals(\sqrt{5})^+}}{\rationals(\sqrt{5})/\rationals}$
  \end{itemize}
\end{prop}
\begin{proof}
  Proof given in appendix \ref{proof:cyclo5}.
\end{proof}

\noindent
Let us now present a second explicit example of a Galois CPM theory.
The cyclotomic extension $\rationals\subset \rationals(\zeta_7)$, where $\zeta_7$ is a primitive seventh root of unity, is our first example of an extension where the decoherence structures do not just form a linear tower.
This is because the Galois group has two non-trivial subgroups with trivial intersection.
The Galois group of $\rationals\subset \rationals(\zeta_7)$ is $\Gamma:=\langle \tau\mid \tau^6=\id{} \rangle \simeq C_6$, where $\tau::\zeta_7\mapsto \zeta_7^3$.
There are two non-trivial subgroups $\Lambda_1 := \langle \tau^3\rangle \simeq C_2$ and $\Lambda_2 := \langle \tau^2\rangle \simeq C_3$.
The subgroups give rise to two intermediate fields $\textrm{Fix}(\Lambda_1) = \rationals(\zeta_7+\zeta_7^6)$ and $\textrm{Fix}(\Lambda_2)=\rationals(\sqrt{-7})$, both of which are Galois extensions of $\rationals$ (note that $\Gamma$ is abelian).

We have the following equalities of closures of field norms:
\begin{align*}
  \clN{\rationals(\zeta_7)}{\rationals} & = \rationals^+ & \clN{\rationals(\zeta_7)}{\rationals(\sqrt{-7})} & = \rationals(\sqrt{-7}) \\
  \clN{\rationals(\zeta_7)}{\rationals(\zeta_7+\zeta_7^6)} & = \rationals(\zeta_7+\zeta_7^6)^+ & \clN{\rationals(\sqrt{-7})}{\rationals} & = \rationals^+
\end{align*}

The Galois CPM category induced by this extension has states of the following form:
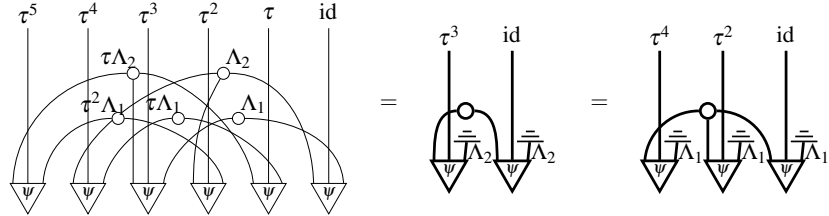
\begin{equation*}
  \tikzfigscale{0.8}{figs/seventhcyclo_state}
\end{equation*}
where the cosets that give rise to each discarding map have been indicated and the automorphisms on $\psi$ have been suppressed for readability.
There are three decoherence maps, the partial decoherences to the intermediate theories and the full decoherence down to the base field:
\begin{equation*}
  \tikzfigscale{0.8}{figs/seventhcyclo_dec1}
\end{equation*}
\begin{equation*}
  \tikzfigscale{0.8}{figs/seventhcyclo_dec2}
\end{equation*}
\begin{equation*}
  \tikzfigscale{0.8}{figs/seventhcyclo_dec3}
\end{equation*}

\begin{prop}\label{prop:cyclo7}
  Define $\quant_{\rationals(\zeta_7)/\rationals}:= \splt{\cpm{\mat{\rationals(\zeta_7)}}{\rationals(\zeta_7)/\rationals}}$ to be the Karoubi envelope for the Galois CPM category from the previous example.
  The full sub-categories of $\quant_{\rationals(\zeta_7)/\rationals}$ spanned by the decoherence maps are characterised as follows:
  \begin{itemize}
    \item $\dec{\rationals(\zeta_7)}{\rationals}$ decoherence maps
    $\longrightarrow$
    equivalent to $\mat{\rationals^+}$
    \item $\dec{\rationals(\zeta_7)}{\rationals(\sqrt{-7})}$ decoherence maps
    $\longrightarrow$
    equivalent to $\cpm{\mat{\rationals(\sqrt{-7})}}{\rationals(\sqrt{-7})/\rationals}$
    \item $\dec{\rationals(\zeta_7)}{\rationals(\zeta_7+\zeta_7^6)}$ decoherence maps
    $\longrightarrow$
    equivalent to $\cpm{\mat{\rationals(\zeta_7+\zeta_7^6)^+}}{\rationals(\zeta_7+\zeta_7^6)/\rationals}$
  \end{itemize}
\end{prop}
\begin{proof}
  Proof given in appendix \ref{proof:cyclo7}.
\end{proof}

\noindent
The result above gives the following correspondence between fields, groups and Galois CPM constructions, respectively:

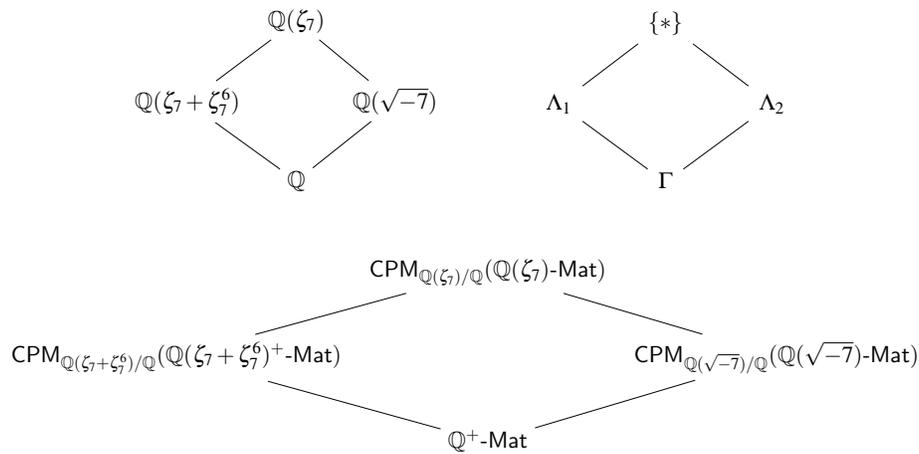
\begin{equation*}
  \adjustbox{scale=0.8}{
    \begin{tikzcd}[cramped, column sep=tiny]
      & \rationals(\zeta_7) \ar[dl,dash] \ar[dr,dash] & \\
      \rationals(\zeta_7+\zeta_7^6)  \ar[dr,dash] & & \rationals(\sqrt{-7}) \ar[dl,dash] \\
      & \rationals &
    \end{tikzcd}
    \qquad \qquad
    \begin{tikzcd}
      & \{*\} \ar[dl,dash] \ar[dr,dash] & \\
      \Lambda_1 \ar[dr,dash] & & \Lambda_2 \ar[dl,dash] \\
      & \Gamma &
    \end{tikzcd}
  }
\end{equation*}

\begin{equation*}
  \adjustbox{scale=0.8}{
    \begin{tikzcd}[cramped, column sep=tiny]
      & \cpm{\mat{\rationals(\zeta_7)}}{\rationals(\zeta_7)/\rationals} \ar[dl,dash] \ar[dr,dash] & \\
      \cpm{\mat{\rationals(\zeta_7+\zeta_7^6)^+}}{\rationals(\zeta_7+\zeta_7^6)/\rationals} \ar[dr,dash] & & \cpm{\mat{\rationals(\sqrt{-7})}}{\rationals(\sqrt{-7})/\rationals} \ar[dl,dash]\\
      & \mat{\rationals^+} &
    \end{tikzcd}
  }
\end{equation*}
In the third diagram, the categories are ``connected'' by the decoherence maps.

\begin{remark}
  Similar proof methods to those of propositions \ref{prop:cyclo5} and \ref{prop:cyclo7} will work for a much larger class of field extensions.
  In particular, for Dedekind groups where all subgroups are normal, all intermediate fields are Galois and many of the results constraining closures of norms will be useful.
  In such a case, the second isomorphism theorem will also come into play allowing one to make similar arguments about morphisms with discarding maps arising from different subgroups.
\end{remark}

\section{Discussion}
In this article we have presented an infinite family of probabilistic theories induced by Galois extensions.
We have seen that these categories contain interesting decoherence structures, in correspondence with subgroups of the Galois group, producing intricate towers of CPM categories with progressively reduced degrees of freedom.
The fundamental theorem of Galois theory acts to provide the bridge between these subgroups and intermediate fields, from which we were able to constrain the scalars that each intermediate theory is over.
Totally imaginary extensions behave especially well: decoherences to other totally imaginary fields are surjective (on the field) while decoherences to formally real intermediate Galois extensions hit only the the totally positive elements.
CM-fields are particularly physically relevant because the scalars of the theory are constrained to be strictly positive and thus have an immediate and standard operational interpretation.
The case of cyclotomic extensions is rich enough to produce non-trivial decoherence towers while being relatively simple fields to work with - they are abelian and constructible by adjoining roots of unity.

More generally, it is the case that the folding of a CPM construction is compositional.
In essence, this is because each folding produces a category with morphisms essentially invariant under the action of the group used to generate it.
Thus, one can factorise a folding into a series of lower-order foldings by subgroups and transversals of the overall group.
Of course in the case of normal subgroups (which is particularly relevant for Galois extensions) this is equivalent to foldings by subgroups and quotient groups.

\subsubsection*{Acknowledgements}
JH wishes to thank Matt Wilson for helpful discussions and acknowledges support from University College London and the EPSRC [grant number EP/L015242/1].
SG wishes to acknowledge support from Hashberg Ltd for this research.

\bibliographystyle{eptcs}
\bibliography{bibliography}

\appendix

\section{A Gentle Introduction to the Classical Galois Theory of Fields}\label{sec:galois}

The core of this article concerns itself with a particular case of CPM categories induced by Galois extensions, particularly finite degree extensions of number fields.
For sake of completeness, we include here a summary of the required tools, techniques and results from the classical Galois theory of fields; for a more detailed discussion, we refer the reader to standard texts \cite{rotman_algebra, bourbaki_algebra}, although any good undergraduate course on the subject would suffice for the core concepts.
This section can be safely skipped by those readers who are already familiar with this most beautiful area of mathematics.

Galois theory concerns itself with understanding field extensions by studying a certain group of field automorphisms known as the \emph{Galois group}.
A field extension $k\subset K$ is simply another way of saying that $k$ is a subfield of $K$ and we will often refer to $k$ as the \emph{base field} and $K$ as the \emph{extension field}.
$K$ always forms a vector space over $k$ and we refer to the dimension of this vector space as the \emph{degree} of the extension, denoting it by $[K:k]$.
An extension is \emph{finite} if the degree is finite; extensions of degree 2 are in particularly called \emph{quadratic}.

\begin{example}
  Here are some examples of field extensions:
  \begin{itemize}
    \item $\reals\subset\complex$ is a quadratic field extension: all $z\in\complex$ can be written in the form $z=a+ib$ for $a,b\in \reals$, so that $1,i$ forms a basis of $\complex$ over $\reals$.
    \item $\rationals\subset \rationals(\zeta_5)$, where $\zeta_5$ is a primitive fifth root of unity, is a fourth degree extension: since roots of unity sum to 0, every element $z\in\rationals(\zeta_5)$ can be written in the form $z=a+b\zeta_5+c\zeta_5^2+d\zeta_5^3$ for some $a,b,c,d\in\rationals$.
    \item $\rationals\subset\rationals(\pi)$ is an infinite degree extension.
  \end{itemize}
\end{example}

\noindent
An element $a\in K$ is \emph{algebraic} over $k$ if there exists a polynomial with coefficients from $k$ that has $a$ as a root.
In other words, there is a polynomial $p\in k[x]$, the ring of polynomials over $k$, such that $p(a)=0$.
A field extension $k\subset K$ is algebraic if every element of $K$ is algebraic over the base field $k$.
In particular, all finite degree extensions are algebraic because for any $a\in K$ there must be some linear dependence between $1,a,a^2,\dots,a^m$ as $m$ grows sufficiently large---or else $K$ cannot be a finite dimensional vector space over $k$---giving a polynomial with $a$ as its root.
An element $a\in K$ which is not algebraic over $k$ is known as \emph{transcendental}; extensions which are not algebraic are also called transcendental.

\begin{example}
  $\reals\subset\complex$ and $\rationals\subset\rationals(\zeta_5)$ are algebraic extensions while $\rationals\subset\rationals(\pi)$ is transcendental.
  On the other hand, $\reals \subset \reals(i\sqrt{\pi})$ is an algebraic extension because $i\sqrt{\pi}$ is a root of $x^2+\pi$.
\end{example}

\noindent
A polynomial $p\in k[x]$ is \emph{irreducible} if $p$ is not a constant polynomial (that is, $\deg{p}>0$) and whenever $p=qr$ for $q,r\in k[x]$ then one of $q$ or $r$ must be a constant polynomial.
An irreducible polynomial is said to \emph{split} over $K$ if it factorises into linear factors over $K$.
In general, it is possible that a polynomial could have a root in $K$ but not all of its roots in $K$: the polynomial would factor, but not split.
An algebraic field extension $k\subset K$ is \emph{normal} if every irreducible polynomial over $k$ which has at least one root in $K$ splits, i.e. has all of its roots in $K$.

\begin{example}
  The polynomial $x^3-2$ has one of its roots in the field $\rationals(\sqrt[3]{2})$ but not its other two.
  Thus this polynomial does not split and this field is not normal.
  Conversely, the extension $\rationals\subset\rationals(\sqrt[3]{2},\omega)$ where $\omega$ is a primitive third root of unity is normal and the polynomial $x^3-2$ now splits.
\end{example}

\noindent
So far we have implicitly viewed extensions such as $\rationals\subset\rationals(\zeta_5)$ as already embedded into some algebraic closure, in this case $\complex$.
When constructing the field $\rationals(\zeta_5)$, we adjoin $\zeta_5$ to $\rationals$ and by knowing which polynomial $\zeta_5$ is a root of we are able to give a general expression for an element of the field.
There is another way of looking at this: we could start with a polynomial and try to construct a field from it directly.
In particular, if $R$ is a commutative ring and $I\subset R$ is an ideal, then the quotient $R/I$ is a field if and only if $I$ is maximal.
One can show that the ideal $(f)$ generated by a polynomial $f\in k[x]$ is maximal if and only if $f$ is irreducible, in which case we can take the quotient $k[x]/(f)$ and form a proper field.
If $a$ is a root of $f$, then there is an isomorphism $k[x]/(f) \simeq k(a)$: $a$ is algebraic over $k$ and $f$ is the \emph{minimal polynomial} of $a$ over $k$.

At this point we are almost in a position to define Galois extensions, but there is a technical pathology which we must rule out: if an irreducible polynomial over $k$ splits in $K$, is it the always case that all of its roots are distinct?
We will not delve too far into the specifics, other than to say that this is always the case for algebraic extensions of finite fields and fields of characteristic zero.
Many extensions---in particular, all those considered in this work---have the property of being \emph{separable}, which rules out the aforementioned pathology: an irreducible polynomial $f\in k[x]$ is called separable if all of its roots are distinct in some extension field, an element $a\in K$ is separable if its minimal polynomial is separable, and an extension is separable if all its elements are separable.

A finite degree field extension $k\subset K$ is \emph{Galois} if it is normal and separable.
It can be shown that this is equivalent to $K$ being the \emph{splitting field} of a separable polynomial $p\in k[x]$, that is $K$ is the minimal degree extension of $k$ such that $p$ splits.

\begin{example}
  The extension $\reals\subset\complex$ is Galois extension, because it is the splitting field of $x^2+1$.
  Similarly, the extension $\rationals\subset\rationals(\zeta_5)$ is Galois, because it is the splitting field of $1+x+x^2+x^3+x^4$.
  The extension $\rationals\subset\rationals(\sqrt[3]{2})$, on the other hand, cannot be Galois, because it is not normal.
  But we have already seen that we get a normal extension $\rationals\subset\rationals(\sqrt[3]{2},\omega)$ if we add $\omega$: this is indeed Galois, because it is the splitting field of $x^3-2$.
\end{example}

Galois extensions have attracted a lot of attention in the study of number fields because of a key property, known as the Fundamental Theorem of Galois Theory.
To understand it, we first need to introduce a few more notions.
If $k \subset L$ and $k \subset K$ are both extensions of $k$, a \emph{$k$-homomorphism} $\tau:L\morph{}K$ is a field homomorphism $L\morph{}K$ which fixes $k$, i.e. one such that $\tau(a)=a$ for all $a\in k$.
The $k$-automorphisms $K\morph{} K$ for a field extension $k\subset K$, i.e. the field automorphisms which fix the base field, always form a group $\aut{K/k}$: for a Galois extension, this is known as the \emph{Galois group} and is written $\gal{K}{k}$.
Importantly, the order of $\gal{K}{k}$ coincides with the degree of the extension, i.e. $|\gal{K}{k}|=[K:k]$.

\begin{thm}[Fundamental Theorem of Galois Theory]
  If $k \subset K$ is a Galois extension, then there is a bijection between subgroups $H \leq \gal{K}{k}$ and intermediate fields $k \subseteq F \subseteq K$.
  Each subgroup $H\leq \gal{K}{k}$ is sent to the field fixed by all elements of $H$:
  \begin{equation*}
    H\mapsto \textrm{Fix}(H) := \{a\in K : \tau(a)=a, \ \forall \tau\in H\}
  \end{equation*}
  Conversely, each intermediate field $k\subseteq F \subseteq K$ is sent to the group of $F$-automorphisms of $K$:
  \begin{equation*}
    F\mapsto \gal{K}{F}:= \{\tau\in \gal{K}{k} : \tau(a)=a, \ \forall a\in F\}
  \end{equation*}
\end{thm}

\begin{example}\label{ex:cyclo1}
  The extension $\reals\subset \complex$ has Galois group $\{\id{},\conj{}\}\simeq C_2$ generated by complex conjugation.
  The extension $\rationals\subset\rationals(\zeta_5)$ has Galois group $\langle \sigma \mid \sigma^4=\id{} \rangle \simeq C_4$ where $\sigma :: \zeta_5\mapsto\zeta_5^2$.
  There is only one non-trivial subgroup, $\langle\sigma^2\rangle\simeq C_2$:
  one can check that $\sigma^2::\zeta_5\leftrightarrow\zeta_5^4, \zeta_5^2\leftrightarrow\zeta_5^3$ fixes $\zeta_5+\zeta_5^4 = (1+\sqrt{5})/2$ and $\zeta_5^2+\zeta_5^3 = (1-\sqrt{5})/2$.
  Therefore, $\textrm{Fix}(\langle\sigma^2\rangle) = \rationals(\sqrt{5})$.
\end{example}

\noindent
For a Galois extension $k\subset K$ with intermediate field $k\subset F\subset K$, it is always the case that $F\subset K$ is a Galois extension.
A corollary of the Fundamental Theorem of Galois Theory also tells us that $k\subset F$ is Galois if and only if $\Lambda:=\gal{K}{F}$ is normal in $\Gamma:=\gal{K}{k}$.
As a consequence, $\gal{F}{k}$ isomorphic to the quotient $\Gamma/\Lambda$.

\begin{example}\label{ex:cyclo2}
  The extension $\rationals\subset\rationals(\zeta_5)$ has an abelian Galois group.
  So the subgroup $$\Lambda:=\gal{\rationals(\zeta_5)}{\rationals(\sqrt{5})} = \langle\sigma^2\rangle\simeq C_2$$ is automatically normal in $\Gamma:=\gal{\rationals(\zeta_5)}{\rationals} \simeq C_4$.
  As a consequence, $\gal{\rationals(\sqrt{5})}{\rationals}\simeq C_4/C_2 \simeq C_2$ and can be explicitly characterised by restricting the automorphisms of $\Gamma$ to the intermediate field $\rationals(\sqrt{5})$:
  \begin{equation*}
    \gal{\rationals(\sqrt{5})}{\rationals} = \{\id{},\sigma\mid_{\rationals(\sqrt{5})}=:\tau \}
  \end{equation*}
  where $\tau::\sqrt{5}\mapsto-\sqrt{5}$ as expected.
\end{example}

The Fundamental Theorem of Galois theory is one of the two notions from Galois theory that will play a major role in this work; the other is the field norm.
All finite extensions $k\subset K$ admit a multiplicative map $N_{K/k}:K\morph{}k$, known as the \emph{field norm}, which sends elements of the extension back to the base field.
To understand how the field norm is defined, note that every element $a\in K$ induces a map $m_a::x\mapsto ax$ by left multiplication.
Because $K$ is a finite-dimensional vector space over $k$, this map has a matrix representation: the field norm of $a$ is defined to be the determinant of this matrix, that is $N_{K/k}(a) := \det(m_a)$.
If $k\subset E$ is Galois, the field norm can be written explicitly as follows:
\begin{equation}\label{eq:norm_sep}
  N_{K/k}(a) = \prod_{\sigma\in \gal{K}{k}} \sigma(a)
\end{equation}
More generally, if $k \subset E$ is a finite separable extension, then the field norm can be written explicitly as follows:
\begin{equation}\label{eq:norm}
  N_{E/k}(a) = \prod_{\sigma\in T} \sigma(a)
\end{equation}
where $T$ is a left transversal of $\gal{\hat{E}}{E}$ in $\gal{\hat{E}}{k}$ \cite{rotman_algebra} and $\hat{E}$ is the normal closure of $E$---the separable field extension of $E$ of smallest degree that is normal (and hence also Galois).

The field norm is a group homomorphism for the multiplicative groups of $K$ and $k$, $N_{K/k}:K^\times \morph{} k^\times$, meaning expressions like $N_{K/k}(ab) = N_{K/k}(a)N_{K/k}(b)$ hold.
Additionally, field norms behave well with towers of extensions, factorising via the intermediate fields.
That is, if we have a tower of field extensions of finite degree $k\subset K \subset L$, then $N_{L/k} = N_{K/k}\circ N_{L/K}$.

\subsection{Number Fields}\label{app:numfields}
We are particularly interested in \emph{number fields}, algebraic extensions of $\rationals$, in this work so we include a compilation of useful results and definitions about them.
From now on when ``field'' is used unqualified, we mean a number field.

A field $K$ is \emph{ordered} if there exists a subset $P\subset K$ which is closed under addition and multiplication, with $K$ equal to the disjoint union $P\sqcup \{0\} \sqcup -P$ where $-P:= \{-p : p\in P\}$.
In such an ordered field one writes $a>b$ if and only if $a-b\in P$.
A field $K$ is \emph{formally real} if -1 is not a sum of squares in $K$ which is equivalent to $K$ being ordered \cite{jacobson_algebra}.
There is a bijection between orderings of $K$ and embeddings (field homomorphisms) of $K$ into its real closure (one can just think of $\reals$ for the fields in this work) and for a Galois number field the embeddings are equivalent to considering the $\rationals$-automorphisms contained in the Galois group.

An element $a\in K$ is \emph{totally positive} if $a>0$ for all orderings of $K$, or equivalently if $\sigma(a)>0$ for all real embeddings $\sigma$ of $K$.
We write $K^+$ for the set of totally positive elements of $K$, which forms a semiring if we additionally include 0.
The semiring of totally positive elements actually has the structure of a \emph{semifield} where every non-zero element has a multiplicative inverse.
If $K$ has no orderings then it is vacuously true that any element is positive for all orderings and we say that all elements of $K$ are totally positive.
It is the case that total positivity is preserved under field norms.

A number field is called \emph{totally real} if all embeddings into the complex numbers lie within the real numbers.
If, on the other hand, no embeddings lie within the real numbers, then the extension is known as \emph{totally imaginary} (or sometimes as \emph{totally complex}).
All Galois number fields are either totally real or totally imaginary and for them, being totally real is equivalent to being formally real and being totally imaginary is equivalent to not being formally real.

The well-known Waring's problem asks whether for each $d\in \naturals$, every natural number is the sum of some finite number $n\in \naturals$ of naturals raised to the $d^\textrm{th}$ power, and was proven by Hilbert in 1909.
The result implies that the same is true for the rationals $\rationals$.

One can ask a similar question of a general field.
If $K$ is a field then we say that \emph{Waring's problem of exponent} $d$ holds if every totally positive element $a\in K$ can be written as a finite sum of $d^\textrm{th}$ powers of totally positive elements of $K$.
That is:
\begin{equation*}
    a = \sum_{i=1}^n a_i^d \qquad a_i\in K
\end{equation*}
where the $a_i$ are all totally positive and $n$ is bounded above by some finite $g(K,d)$ dependent only on $K$ and $d$.
Ellison \cite{ellison1,ellison2} reduces this problem to being able to write all totally positive elements of $K$ as a finite sum of squares alongside a constraint about the density of $d^\textrm{th}$ powers in $K$.
By a classical result of Siegel \cite{siegel} the former of these is possible for number fields: every totally positive element of $K$ is a sum of at most four squares in $K$.

For us, the outcome of this discussion is the following useful result:
\begin{thm}[Waring's Problem for Fields \cite{ellison1,ellison2}]\label{thm:waring}
  If either of the following hold:
  \begin{enumerate}
    \item $K$ is a non-real field of characteristic 0
    \item $K$ is formally real, every totally positive element of $K$ can be written as a sum of at most $s$ squares for some $s$, and $d^\textrm{th}$ powers are suitably dense in $K$
  \end{enumerate}
  then Waring's problem holds for all exponents.
  In particular, if $K$ is a number field then Waring's problem holds for all exponents.
\end{thm}

\section{Further Examples}

\subsection{Quadratic Fields}

A quadratic extension $\rationals\subset K$ is a degree two extension of the rationals, i.e. $[K:\rationals]=2$.
Any quadratic field is isomorphic to one of the form $\rationals(\sqrt{d})$ for $d$ square-free.
If $d>0$ then we have a real quadratic field, while $d<0$ gives an imaginary quadratic field.
Imaginary quadratic fields are clearly CM, and constitute the motivating examples for the theory of CM-fields.

Quadratic fields are Galois extensions of $\rationals$ (they are the splitting fields of $x^2-d$) with Galois group isomorphic to $C_2$, generated by the map $\sigma::\sqrt{d}\mapsto -\sqrt{d}$.
Since standard quantum theory has underlying $C_2$ folding symmetry, the CPM categories induced by quadratic fields look a lot like standard quantum theory.

In the case of imaginary quadratic fields, the similarities are substantial.
Writing $\sqrt{d} = i\sqrt{c}$, the field norm $N(x+\sqrt{d}y) = x^2 + cy^2 \geq 0$ is elliptic and non-negative (consistently with our previous results on folding of scalars in CM-fields).
The scalars form the semiring $\rationals^+$ and the phases form a subgroup of the standard quantum phases.
By Hilbert's Theorem 90, these phases take the form $\sigma(b)/b$ for some $b\in \rationals(\sqrt{d})$.

In the case of real quadratic fields, there are instead substantial differences from quantum theory.
The scalars become the entire field $\rationals$ and the norm $N(x+\sqrt{d}y) = x^2 - dy^2$ is hyperbolic.
Hilbert 90 allows for the same description of the phases and the theory has similarities to hyperbolic quantum theory \cite{gogioso_fantastic}: the only multiplicative characters are the real ones, so that hidden subgroup problems \cite{vicary_topology, gogioso_hiddensubgroup} can only be efficiently solved for the groups $\mathbb{Z}_2^n$.

\subsection{Finite Fields}

Galois CPM categories induced by finite fields provide simple and nice examples of the structures we have seen so far.
For any power of a prime $q=p^n$ there exists a unique finite field $\finites{q}$ of order $q$.
The non-zero elements $\finites{q}^\times$ form a cyclic group of order $q-1$, generated by some element $a$.
Extensions of the form $\finites{q}\subset \finites{q^m}$ are always Galois with cyclic Galois group generated by the Frobenius automorphism $\phi_p:: t\mapsto t^p$ for $t\in\finites{q^m}$.

Of particular interest to us is the fact that the field norm is surjective.
Indeed, by taking $a$ to generate $\finites{q^m}^\times$, one always has that $N_{\finites{q^m}/\finites{q}}(a) = a^{1+q+\dots+q^{m-1}} = a^{(q^m-1)/(q-1)}$, immediately implying that $N(a)$ has multiplicative order $q-1$; therefore $N(a)$ generates $\finites{q}$.
Because the field norm is surjective, the endomorphisms of the unit object $\textrm{End}(I)$ of $\cpm{\mat{\finites{q^m}}}{\finites{q^m}/\finites{q}}$ forms a field isomorphic to $\finites{q}$, and all intermediate theories spanned by decoherence maps in $\quant_{\finites{q^m}/\finites{q}}$ are Galois CPM categories for finite fields, not just categories over sub-semirings.

\begin{prop}\label{prop:finitefields}
  Let $\quant_{\finites{q^m}/\finites{q}}$ be the subcategory of the Karoubi envelope of $\cpm{\mat{\finites{q^m}}}{\finites{q^m}/\finites{q}}$ spanned by the decoherence maps induced by the subgroups of the Galois group.
  Then for any subgroup $\Lambda\leq \Gamma$, the subcategory of $\quant_{\finites{q^m}/\finites{q}}$ spanned by the decoherence maps $\dec{}{\Lambda}$ is equivalent to the category $\cpm{\mat{\finites{q^l}}}{\finites{q^l}/\finites{q}}$ where $\finites{q^l} = \textrm{Fix}(\Lambda)$.
\end{prop}
\begin{proof}
  Proof given in appendix \ref{proof:finitefields}.
\end{proof}

\section{CPM Categories for Separable Extensions}\label{sec:separable}
Here we sketch a generalisation of the construction presented in the main article, where we have an extension $k\subset E$ which is separable but not necessarily normal.
As the full story is not complete yet, we relegate this discussion to an appendix, but we hope it goes to show that the constructions developed here can be applied to a much larger class of field extensions than discussed in the main article.

As noted in appendix \ref{sec:galois}, a separable extension still has a concise expression for the field norm $N_{E/k}$, but we are forced to work with the normal closure $\hat{E}$ of $E$.
Both $k\subset\hat{E}$ and $E\subset\hat{E}$ are then Galois extensions, and the norm $N_{E/k}$ is given by a product over a transversal of $\gal{\hat{E}}{E}$ in $\gal{\hat{E}}{k}$ (see equation \eqref{eq:norm_sep}).

The generalisation of the CPM construction to group transversals outlined in section \ref{sec:cpmcats} gives the necessary machinery to treat the case of separable extensions.
One way of producing an interesting CPM category (with suitably constrained scalars) from $E$ is to ``upgrade'' $\mat{E}$ to $\mat{\hat{E}}$, which has a canonical group action $\phi$ by its Galois group $\Gamma:=\gal{\hat{E}}{k}$.
Upon picking a left transversal $T$ of $\Lambda:=\gal{\hat{E}}{E}$ in $\Gamma$, one can then take the folding functor $\fld{\tau}:\mat{E} \morph{} \mat{\hat{E}}$, where $\mat{E}$ is equivalent to the subcategory of the equivariant category $\mat{\hat{E}}_\Lambda$ spanned by the identity isomorphisms $\eta_A^g := \id{A}$ (because $\phi_g$ is the identity on objects for all $g \in \Lambda$).

One consequence of this generalisation is that we are now able to consider Galois CPM categories induced by extensions with Galois groups which are not \emph{Dedekind}: that is, Galois groups with subgroups which are not normal.
For such an extension there exist intermediate fields which are not Galois over the base field (they are in bijection with the non-normal subgroups) and complete foldings are not enough to deal with them in a rigorous way.

As an example, consider the extension $\rationals\subset \rationals(\alpha,\omega)$, where $\omega$ is a primitive third root of unity and $\alpha^3=2$.
This is the splitting field of $x^3-2$ over $\rationals$, and is therefore Galois.
It has Galois group $\Gamma \simeq S_3$ generated by the automorphisms $\sigma::\omega\mapsto\omega^2$ and $\tau::\alpha\mapsto\alpha\omega$, with the following lattice of subgroups:
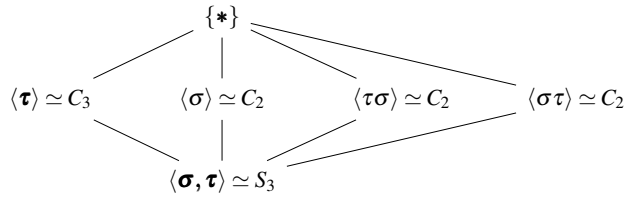
\begin{equation*}
  \adjustbox{scale=0.8}{
    \begin{tikzcd}
      &  \bm{\{* \}} \ar[dl,dash] \ar[d,dash] \ar[dr,dash] \ar[drr,dash] & & \\
      \bm{\langle\tau\rangle} \simeq C_3 \ar[dr,dash] & \langle\sigma\rangle \simeq C_2 \ar[d,dash] & \langle\tau\sigma\rangle \simeq C_2 \ar[dl,dash] & \langle\sigma\tau\rangle \simeq C_2 \ar[dll,dash] \\
      & \bm{\langle\sigma,\tau\rangle} \simeq S_3  & &
    \end{tikzcd}
  }
\end{equation*}

The normal subgroups of $\Gamma$ are shown here in bold font.
This lattice is in bijection with the following lattice of intermediate fields:
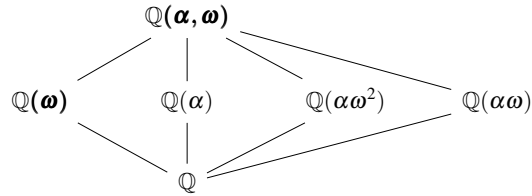
\begin{equation*}
  \adjustbox{scale=0.8}{
    \begin{tikzcd}
      &  \bm{\rationals(\alpha,\omega)} \ar[dl,dash] \ar[d,dash] \ar[dr,dash] \ar[drr,dash] & & \\
      \bm{\rationals(\omega)} \ar[dr,dash] & \rationals(\alpha) \ar[d,dash] & \rationals(\alpha\omega^2) \ar[dl,dash] & \rationals(\alpha\omega) \ar[dll,dash] \\
      & \bm{\rationals} & &
    \end{tikzcd}
  }
\end{equation*}
The Galois extensions of $\rationals$ are shown here in bold font.
With the generalisation of the CPM construction given in this section, we are now able to consider decoherences to the non-normal intermediate fields, such as $\rationals(\alpha)$.
The folding of $\rationals(\alpha,\omega)$ over $\rationals(\alpha)$ is straightforward, since this extension is Galois.
The folding of $\rationals(\alpha)$ over $\rationals$ is more tricky, and requires picking a left transversal of $\langle\sigma\rangle$ in $\Gamma$; for instance, one can pick $T:= \{\id{},\tau,\tau^{-1}\}$.
These foldings mimic the form of the field norms and act to suitably constrain the scalars of the theories.
\begin{equation*}
  N_{\rationals(\alpha,\omega)/\rationals(\alpha)}(a) = a\sigma(a) \qquad N_{\rationals(\alpha)/\rationals}(a) = a\tau(a)\tau^{-1}(a)
\end{equation*}
Although it was possible to pick $T$ to be a subgroup of $\Gamma$ this need not be the case in general.
A transversal can always be given the structure of a quasigroup, and if the transversal contains the identity then the algebraic structure is stronger and forms a loop.
These structures are not associative, making them problematic to study internally to a category which is why we take the route of describing the foldings directly at the level of the transversal.
Nevertheless, there are many occasions when a particular choice of transversal does form a group - yet one must take care since this group is not, even in the case where $\Lambda$ is normal in $\Gamma$, necessarily a subgroup of $\Gamma$ (for instance consider the quaternion group $Q_8$ which has centre $Z(Q_8)\simeq C_2$ and quotient $Q_8/Z(Q_8)\simeq C_2\times C_2$ which is not isomorphic to any subgroup of $Q_8$).
Finally, we can write down the decoherence maps in all their glory:
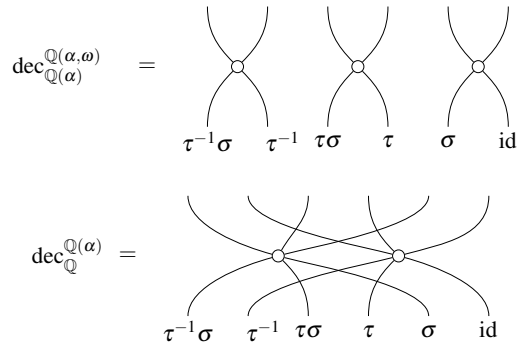
\begin{equation*}
  \tikzfigscale{0.8}{figs/s3dec1}
\end{equation*}
\begin{equation*}
  \tikzfigscale{0.8}{figs/s3dec2}
\end{equation*}

\section{Proofs}

\subsection{Proof of Proposition \ref{prop:equivariantmonoidal}}\label{proof:equivariantmonoidal}
\begin{proof}
  The monoidal structure $\otimes$ of $\cat{C}$ induces a monoidal structure $\boxtimes$ on $\cat{C}_G$.
  Strictness of $\phi_g$ implies that $\eta_A^g\otimes\eta_B^g$ have the correct type.
  Everything else quickly follows.
\end{proof}

\subsection{Proof of Proposition \ref{prop:naturaliser}}\label{proof:naturaliser}
\begin{proof}
$\textrm{Nat}((\eta^h)_{h \in H})$ is the largest subcategory of $\cat{C}$ such that all $\eta^h:1\Rightarrow\phi_h$ are natural isomorphisms.
It is a wide subcategory containing only the arrows $f$ which make diagram \ref{diag:equivariantmorphs} commute.
The equivalence to $\hat{\cat{C}}_{H,\eta}$ follows immediately by forgetting the families $(\eta_a^h)_{h \in H}$ involved in the objects, that is sending $(A,(\eta_A^h)_{h \in H})$ to $A$ in $\textrm{Nat}((\eta^h)_{h \in H})$.
\end{proof}

\subsection{Proof of Proposition \ref{prop:autofunctors_folded}}\label{proof:autofunctors_folded}
\begin{proof}
  The original proof from \cite{gogioso_cpm}, valid for complete folding functors, requires slight tweaking for our generalised setting.
  Any $g\in G$ acts on left transversals of $H \leq G$ by sending a left transversal $T$ to another left transversal $T'$, obtained by permuting the left cosets and altering the choice of element in each coset.
  There is an isomorphism $\sigma_A^g$, given by a suitable composition of the symmetry isomorphisms $\sigma$ from the symmetric monoidal structure, which arranges the cosets into their original order.
  There is also an isomorphism $\rho_A^g$, given by a monoidal product of the isomorphisms $\eta_A^h$, which acts to recover the original choices in the transversal.
  Taken together, $\sigma_A^g$ and $\rho_A^g$ compose to an isomorphism between the folded object $\fld{\tau}A$ and the action of $G$ on the indices in the monoidal product.
  \begin{equation*}
    \phi_g \fld{\tau} A
    =
    \phi_g\bigotimes_{t\in T}\phi_t \iota A
    =
    \bigotimes_{t\in T} \phi_{gt}\iota A
    =
    \bigotimes_{t'\in T'}\phi_{t'}\iota A \morph{\sigma_A^g} \bigotimes_{t\in T} \phi_{t h_t}\iota A \morph{\rho_A^g := \bigotimes\limits_{t\in T} \phi_t (\eta^{h_t}_A)^{-1} } \bigotimes_{t\in T} \phi_t \iota A
    =
    \fld{\tau} A
  \end{equation*}
  where $h_t$ in an element of $H$ for each $t\in T$.
  We then have:
  \begin{align*}
    \rho_A^g\circ\sigma_A^g\circ(\phi_g \fld{\tau}f) \circ {\left(\sigma_A^g\right)}^{-1} \circ {\left(\rho_A^g\right)}^{-1} & = \rho_A^g\circ\sigma_A^g\circ \bigotimes_{t\in T} \phi_{gt}\iota f \circ {\left(\sigma_A^g\right)}^{-1} \circ {\left(\rho_A^g\right)}^{-1} \\
    & = \rho_A^g \circ\sigma_A^g\circ \bigotimes_{t'\in T'} \phi_{t'}\iota f \circ {\left(\sigma_A^g\right)}^{-1} \circ {\left(\rho_A^g\right)}^{-1} \\
    & = \rho_A^g \circ \bigotimes_{t\in T} \phi_{th_t}\iota f \circ {\left(\rho_A^g\right)}^{-1} \\
    & = \bigotimes_{t\in T}\phi_t \iota f = \fld{\tau}f
  \end{align*}
  This completes our proof.
\end{proof}

\subsection{Proof of Proposition \ref{prop:foldfactor}}\label{proof:foldfactor}
\begin{proof}
  The $G$-action restricts to an $H$-action, giving a complete folding functor $\fld{H}:\cat{C}\morph{}\FLD{H}(\cat{C})$.
  The category $\FLD{H}(\cat{C})$ is $H$-equivariant with respect to isomorphisms $\sigma^h$ given by a suitable composition of the symmetry isomorphisms from the symmetric monoidal structure; these isomorphisms merely act by ``rearranging'' the order of the tensor factors.
  Therefore, there is an embedding $e:\FLD{H}(\cat{C})\morph{}\hat{\cat{C}}_{H,\eta}$ of the image of $\fld{H}$ into $\hat{\cat{C}}_{H,\eta}$ given by sending $\fld{H}f:\fld{H}A\morph{}\fld{H}B$ to $\fld{H}f:(\fld{H}A,\{\sigma_A^h\})\morph{}(\fld{H}B,\{\sigma_B^h\})$.
  Taking $T$ to be a transversal of $H$ in $G$, we can form the folding functor $\fld{\tau}:\hat{\cat{C}}_{H,\eta}\morph{}\cat{C}$ and observe that the following diagram commutes:
  \begin{equation*}
    \begin{tikzcd}
      \cat{C} \ar[r,"\fld{H}"]  \ar[d,"\fld{G}"']& \FLD{H}(\cat{C}) \ar[d,"e"]\\
      \cat{C} & \hat{\cat{C}}_{H,\eta} \ar[l,"\fld{\tau}"]
    \end{tikzcd}
  \end{equation*}
  This completes our proof.
\end{proof}

\subsection{Proof of Proposition \ref{prop:foldfactor2}}\label{proof:foldfactor2}
\begin{proof}
  We will make use of the following definition:
  \begin{defn}[$G$-functor]
    Let $\cat{C}$ and $\cat{C}'$ be two categories equipped with $G$-actions $\phi$ and $\phi'$ respectively.
    A \emph{$G$-functor} $(f,\sigma):(\cat{C},\phi)\morph{}(\cat{C}',\phi')$ is a pair of a functor $f:\cat{C}\morph{}\cat{C}'$ and a family of natural isomorphisms $\sigma_g:f\phi_g\morph{}\phi_g' f$ for each $g\in G$, such that the following diagram commutes:
    \begin{equation*}
      \begin{tikzcd}
        f\phi_g\phi_h \ar[r,"\sigma_g\phi_h"] \ar[rd,"\sigma_{gh}"']& \phi_g' f\phi_h \ar[d,"\phi_g'\sigma_h"] \\
        & \phi_g'\phi_h' f
      \end{tikzcd}
    \end{equation*}
    If $f$ is an equivalence of categories, then we call this a \emph{$G$-equivalence}.
  \end{defn}
  It is known that there exists an induced action of $G$ on $\cat{C}_H$ given by taking $\psi_g:\cat{C}_H\morph{}\cat{C}_H$ to act on morphisms as $\phi_g$ and on objects as $(A,(\eta_A^h)_{h \in H})\mapsto (\phi_g A, (\phi_g\eta_A^{g^{-1}hg})_{h \in H})$ \cite{beckmann_equivariant}.
  This descends to an action of the quotient by picking a transversal of $H$ in $G$ containing the identity and defining $\hat{\psi}_{gH}:=\psi_g$ where $g$ is the representative of the coset $gH$ in the transversal.
  Any two different choices of transversal $T$ and $T'$ result in different $G/H$-actions $\hat{\psi}$ and $\hat{\psi}'$, but there is a $G/H$-equivalence $(\id{\cat{C}_H},\sigma):(\cat{C}_H,\hat{\psi})\morph{}(\cat{C}_H,\hat{\psi}')$ given by the natural isomorphisms $\sigma_g = \phi_g{\eta^h}$ where $h$ is such that $g' = gh$ for $g$ and $g'$ the representatives of $gH$ in $T$ and $T'$ respectively.
  Thus the action is essentially unique.
  With the $G/H$-action $\hat{\psi}$ one can form a complete folding functor $\fld{G/H}:\hat{\cat{C}}_{H,\eta}\morph{}\hat{\cat{C}}_{H,\eta}$.

  Now, consider the complete folding functor $\fld{G}:\cat{C}\morph{}\cat{C}$.
  As in the previous proposition, this restricts to a folding functor $\fld{H}:\cat{C}\morph{}\FLD{H}(\cat{C})$, and we have the embedding $e:\FLD{H}(\cat{C})\morph{}\hat{\cat{C}}_{H,\eta}$ of the image of $\fld{H}$ into $\hat{\cat{C}}_{H,\eta}$.
  The following diagram then commutes:
  \begin{equation*}
    \begin{tikzcd}
      & \FLD{H}(\cat{C}) \ar[rr,"e"] & & \hat{\cat{C}}_{H,\eta} \ar[rd,"\fld{G/H}"]&\\
      \cat{C} \ar[ur,"\fld{H}"] \ar[rrd,"\fld{G}"'] & & & &\hat{\cat{C}}_{H,\eta} \ar[lld,"\iota"] \\
      & & \cat{C}&&
    \end{tikzcd}
  \end{equation*}
  This concludes our proof.
\end{proof}

\subsection{Proof of Proposition \ref{prop:cpmmonoidal}}\label{proof:cpmmonoidal}
\begin{proof}
  This was originally shown in \cite{gogioso_cpm} for the complete folding case. $\cpm{\cat{C}}{\tau,\Xi}$ is a symmetric monoidal category with a monoidal product induced by that of $\cat{C}$.
  It is easy to check that the original proof generalises to the construction presented here.
\end{proof}

\subsection{Proof of Proposition \ref{prop:cpmdagger}}\label{proof:cpmdagger}
\begin{proof}
  The dagger of $\cat{C}$ extends to the folded category by defining $(f\boxtimes g)^\dag := f^\dag \boxtimes g^\dag$ and $\fld{\tau}(f)^\dag = \fld{\tau}(f^\dag)$.
  This is precisely the dagger of $\FLD{\tau}(\cat{C})$ as a subcategory of $\cat{C}$.

  It is easy to check (cf. \cite{gogioso_cpm}) that any morphism $b$ of $\cpm{\cat{C}}{\tau,\Xi}$ has a normal form, namely $b = (\id{\fld{\tau}B}\boxtimes \xi_E)\circ\fld{\tau}(a)$ for some morphism $a:A\morph{}B\otimes E$ of $\cat{C}$ and some effect $\xi_E\in\Xi_E$.
  Since $\cpm{\cat{C}}{\tau,\Xi}$ is a subcategory of $\cat{C}$, the dagger descends immediately to give $b^\dag = \fld{\tau}(a^\dag)\circ (\id{\fld{\tau}B}\boxtimes \xi_E^\dag)$: all we need to show is that the latter is still a morphism in $\cpm{\cat{C}}{\tau,\Xi}$.
  Using the $\dag$-compact closure of $\cat{C}$ and writing $\text{cup}_A:I\morph{}A^*\otimes A$ for the unit, it follows that:
  \begin{equation}\label{eq:gdag}
    b^\dag = [\id{\fld{\tau}A} \boxtimes \xi_E^*] \circ [\fld{\tau}a^\dag \boxtimes \id{\fld{\tau}E^*}] \circ [\id{\fld{\tau}B} \boxtimes \fld{\tau}\text{cup}_E^*]
  \end{equation}
  By assumption the conjugation functor is one of the autofunctors $\phi_g$, so $\xi_E^* = \xi_E \circ {\left(\sigma_E^g\right)}^{-1} \circ (\rho_A^g)^{-1}$ and so each constituent map of \eqref{eq:gdag} is in the normal form.
  It is fairly straightforward to check the remaining requirements for this to give a valid dagger structure on $\cpm{\cat{C}}{\tau,\Xi}$.

  The compact closure of $\cat{C}$ then implies that $\cpm{\cat{C}}{\tau,\Xi}$ is also compact closed---its unit and counit given by folding those of $\cat{C}$---and the discussion above is enough to conclude that the unit and counit are daggers of each other, giving us $\dag$-compact closure.
\end{proof}

\subsection{Proof of Proposition \ref{prop:H-environment}}\label{proof:H-environment}
\begin{proof}
  The monoidal product of $\dag$-SCFAs is again a $\dag$-SCFA so the environment structure is closed under this operation.
  The autofunctors act essentially trivially on any effect in the canonical environment structure: if $h \in H$ then $h$ just permutes the legs of each spider and if $h\notin H$ then it acts to also permute the spiders.
  In either case there exists a natural isomorphism given by the symmetric monoidal structure of $\cat{C}$ which acts to undo the permutation and this is all we require because we made the assumption that the isomorphisms $\theta_A^g$ satisfy the commutative diagram \eqref{diag:eta} and act on the $\dag$-SCFA in essentially the same way as the autofunctors $\phi_g$.
\end{proof}

\subsection{Proof of Proposition \ref{prop:pos_in_norm} and Corollary \ref{cor:nonreal}}\label{proof:pos_in_norm}
\begin{proof}
  Say $[K:k]=d$, then for any $a\in k \hookrightarrow K$ we have $\N{K}{k}(a) = a^d$.
  Thus all finite sums $\sum_i a_i^d \in \clN{K}{k}$ for $a_i\in k$.
  By Waring's problem this is all totally positive elements.

  When $k$ is not formally real, every element of $k$ is totally positive so $\clN{K}{k}=k$
\end{proof}

\subsection{Proof of Proposition \ref{prop:imagtoreal}}\label{proof:imagtoreal}
\begin{proof}
  The embeddings of a totally imaginary field always come in pairs, for if $e:K\hookrightarrow\complex$ is an embedding then complex conjugation $J_\complex:\complex\morph{}\complex$ composed with $e$ gives another embedding of $K$.
  Indeed, $J_\complex$ induces an automorphism $J$ of $K$ which acts like complex conjugation on $K$.
  This automorphism is not necessarily independent of the choice of embedding into $\complex$ (and in particular will not commute with the other elements of the Galois group, unless for instance, the field is CM), but nevertheless it forms a subgroup of $\Gamma:=\gal{K}{\rationals}$ isomorphic to $C_2$.

  Now, $k$ is totally real and so must be fixed by $J$ implying that $J\in\Lambda:=\gal{K}{k}$.
  Thus there is at least one embedding of $k$ into $\reals$ where $\N{K}{k}(a)>0$ for all $0\neq a\in K$.

  Fix this embedding and consider any $\sigma\in\gal{k}{\rationals}$. We have:
  \begin{equation*}
    \sigma \N{K}{k}(a) = \prod_{\lambda\in\Lambda}\sigma\lambda(a) = \prod_{\mu\in\sigma\Lambda}\mu(a) = \prod_{\mu\in\Lambda\sigma} \mu(a) = \N{K}{k}(\sigma(a)) > 0
  \end{equation*}
  where we used that fact that $\Lambda$ must be normal in $\Gamma$ and thus left cosets and right cosets coincide.
  So $\N{K}{k}(a)$ is positive for all embeddings of $k$.
  Therefore $\overline{N}_{K/k} \subset k^+$.
  Proposition \ref{prop:pos_in_norm} gives the other containment $k^+ \subset \overline{N}_{K/k}$.
\end{proof}

\subsection{Proof of Proposition \ref{prop:totreal_norm}}\label{proof:totreal_norm}
\begin{proof}
  This is immediate if $[K:\rationals]$ is odd: just note $N(-1)=-1$ and combine with proposition \ref{prop:pos_in_norm}.
  If $[K:\rationals]$ is even then a different argument is needed (which still holds for the odd case).

  By the normal basis theorem we know that there exists some $\alpha\in K$ such that $\{\sigma(\alpha) : \sigma\in\gal{K}{\rationals}\}$ forms a $\rationals$-basis of $K$.
  This means that $\alpha$ is distinct under all Galois automorphisms $\sigma_i$.
  Since $K$ is totally real, there is an ordering on $K$ given by the ordering of $\reals$.

  If there are an odd number of $\sigma_i$ such that $\sigma_i(\alpha)<0$ then $N(\alpha)<0$ and the result follows.

  If there are an even number $2m$ of $\sigma_i$ such that $\sigma_i(\alpha)<0$ then, ignoring the $\sigma_i$ where $\sigma_i(\alpha)>0$, we have an ordering, say $\sigma_1(\alpha)< \dots < \sigma_{2m}(\alpha) < 0$.
  There exists $q\in \rationals$ such that $-\sigma_{2m}(\alpha) < q < -\sigma_{2m-1}(\alpha)$ which implies that $\sigma_1(\alpha+q)<\dots<\sigma_{2m-1}(\alpha+q)<0<\sigma_{2m}(\alpha+q)$.
  Thus $N(\alpha+q)<0$ and the result follows.
\end{proof}

\subsection{Proof of Proposition \ref{prop:cyclo5}}\label{proof:cyclo5}
\begin{proof}
  We will directly construct the functors and show that they are full, faithful and essentially surjective on objects and thus witness the equivalences.

  Start with the decoherences $\dec{\rationals(\zeta_5)}{\rationals}$.
  On objects send $(\fld{\Gamma}n,\dec{\rationals(\zeta_5)}{\rationals})$ to $n$ in $\mat{\rationals^+}$, this is clearly essentially surjective on objects.
  On morphisms $f:(\fld{\Gamma}n,\dec{\rationals(\zeta_5)}{\rationals}) \morph{} (\fld{\Gamma}m,\dec{\rationals(\zeta_5)}{\rationals})$ we do the following:

  \begin{equation*}
    \tikzfigscale{0.8}{figs/dec_equiv_cyclo5}
  \end{equation*}
  which is clearly faithful.
  We are left to show that the functor is full, which follows by the results on the closures of norms.
  The symmetries of the maps are sufficient to show that the elements of the matrices are in $\rationals$ -  they are fixed by the Galois group.
  The fact that $\clN{\rationals(\zeta_5)}{\rationals}=\rationals^+$ shows that it is enough to sum pure maps of the form $\fld{\Gamma}g$ for $g:n\morph{}m$ using the complete discarding maps in order to get every matrix of $\mat{\rationals^+}$.
  That we can get nothing more than this is a consequence of the preservation of totally positive elements under norms.
  A general fully decohered map of $\quant_{\rationals(\zeta_5)/\rationals}$ may contain $\Lambda$-discarding maps but the matix elements of such a map can always be written in the form $\sum_i \N{\rationals(\sqrt{5})}{\rationals} (\alpha_i)$ for $\alpha_i \in \clN{\rationals(\zeta_5)}{\rationals(\sqrt{5})} = \rationals(\sqrt{5})^+$.
  At which point it is enough to note that totally positive elements of $\rationals(\sqrt{5})$ are sent to totally positive elements of $\rationals$ by the norm.

  A similar argument holds for the $\dec{\rationals(\zeta_5)}{\rationals(\sqrt{5})}$ decoherences.
  On objects send $(\fld{\Gamma}n,\dec{\rationals(\zeta_5)}{\rationals})$ to $\fld{\Lambda}n$ and on morphisms do similar to the previous case of combining input or output legs on each spider.
  The symmetries of the maps show that the elements of the matrices live in $\rationals(\sqrt{5})$ and that we can only reach the totally positive ones is a consequence of the closure of the norm.
\end{proof}

\subsection{Proof of Proposition \ref{prop:cyclo7}}\label{proof:cyclo7}
\begin{proof}
  The proof is very similar to proposition \ref{prop:cyclo5} - the construction of the functors is analogous and the arguments about norms hold here too, so we leave the reader to fill in those details.
  There is one major additional point which needs to be considered: what happens when we have a morphism with both $\Lambda_1$ and $\Lambda_2$-discarding maps?
  We must demonstrate that these do not give rise to any additional maps (i.e. a matrix with some non-positive entries).

  For instance consider an entry of a $\Lambda_1$-decohered morphism with a $\Lambda_2$-discarding map and note that the following holds:
  \begin{equation*}
    \tikzfigscale{0.8}{figs/seventhcyclo_crossdec}
  \end{equation*}
  so that such a term is in the image of the expected norm.
  A similar result holds reversing $\Lambda_1$ and $\Lambda_2$ and both are precisely because of the second isomorphism theorem, also known as the diamond theorem.
  This tells us that $\Gamma/\Lambda_1 \simeq \Lambda_2/\{*\} \simeq \Lambda_2$ and so the folding due to $\Lambda_2$ is precisely the same as the quotient folding $\Gamma/\Lambda_1$.
  In other words, the folding ``left over'' after decohering by $\Lambda_1$ is that of $\Lambda_2$.

  We have dealt with the decoherences to the intermediate fields, but there is one more issue which may raise some concern - the full decoherence to $\rationals$ but on a morphism with both $\Lambda_1$ and $\Lambda_2$-discarding maps.
  In fact, by a similar method to the one outlined above one can show that an entry in such a matrix must be an element of both the closure of the norms to both intermediate fields, while of course also being an element of $\rationals$ by the symmetries.
  Thus it must be an element of $\rationals\cap\rationals(\sqrt{-7})\cap\rationals(\zeta_7+\zeta_7^6)^+ = \rationals^+$.
\end{proof}

\subsection{Proof of Proposition \ref{prop:finitefields}}\label{proof:finitefields}
\begin{proof}
  The proof is very similar to propositions \ref{prop:cyclo5} and \ref{prop:cyclo7}.
  By observing that the field norms are surjective for finite fields, so that their images are isomorphic to the entire codomain field, one does not need to be nearly as careful as for the aforementioned propositions.
  The simple symmetry argument will suffice.
\end{proof}

\end{document}

%% file: preamble.tex
\usepackage[utf8]{inputenc}
\usepackage{amsmath}
\usepackage{amsthm}
\usepackage{amssymb}
\usepackage{hyperref}
\usepackage{physics}
\usepackage{tikz}
\usepackage{tikz-cd}
\usepackage{tikzit}
\usepackage{circuitikz}
\usepackage{bm}
\usepackage[T1]{fontenc}
\usepackage{adjustbox}

\input{styles.tikzdefs}
\input{styles.tikzstyles}
\usetikzlibrary{circuits.ee.IEC}

\newcommand{\morph}[1]{\xrightarrow{#1}}
\newcommand{\reals}{\mathbb{R}}
\newcommand{\complex}{\mathbb{C}}
\newcommand{\integers}{\mathbb{Z}}
\newcommand{\naturals}{\mathbb{N}}
\newcommand{\rationals}{\mathbb{Q}}

\newcommand{\finites}[1]{\mathbb{F}_{#1}}

\newcommand{\cat}[1]{\mathcal{#1}}
\newcommand{\cpm}[2]{\mathsf{CPM}_{#2}(#1)}
\newcommand{\fhilb}{\mathsf{FHilb}}
\newcommand{\gal}[2]{\textrm{Gal}(#1/#2)}
\newcommand{\aut}[1]{\textrm{Aut}(#1)}
\newcommand{\fld}[1]{\textrm{fld}_{#1}}
\newcommand{\FLD}[1]{\mathsf{FLD}_{#1}}
\newcommand{\id}[1]{\textrm{id}_{#1}}
\newcommand{\mat}[1]{#1\textrm{-}\mathsf{Mat}}
\newcommand{\dec}[2]{\textrm{dec}^{#1}_{#2}}
\newcommand{\splt}[1]{\mathsf{Split}(#1)}

\newcommand{\conj}[1]{\textrm{conj}_{#1}}
\newcommand{\quant}{\mathsf{Quant}}
\newcommand{\N}[2]{N_{#1 / #2}}
\newcommand{\clN}[2]{\overline{N}_{#1 / #2}}

\newcommand{\Ground}{%
  \mathbin{\text{\begin{tikzpicture}[circuit ee IEC,yscale=0.6,xscale=0.7]
  \draw (0,-2ex) to (0,0.5ex) node[ground,rotate=90,xshift=.75ex] {};
  \end{tikzpicture}}}%
}

\newcommand{\tikzfigscale}[2]{\scalebox{#1}{\input{#2.tikz}}}

\newcommand{\zcolour}{white}

\theoremstyle{definition}
\newtheorem{defn}{Definition}
\theoremstyle{plain}
\newtheorem{prop}{Proposition}
\theoremstyle{plain}

\theoremstyle{plain}
\newtheorem{thm}{Theorem}
\theoremstyle{plain}
\newtheorem{cor}{Corollary}[prop]
\theoremstyle{remark}
\newtheorem*{remark}{Remark}
\theoremstyle{definition}
\newtheorem{example}{Example}

%% file: styles.tikzstyles
\tikzstyle{discarding}=[fill=white, draw=black, shape=circle, style=upground]
\tikzstyle{smalldiscarding}=[fill=white, draw=black, style=upground, scale=0.6]
\tikzstyle{backdiscard}=[fill=white, draw=black, shape=circle, style=downground, scale=0.5]
\tikzstyle{smallbackdiscard}=[fill=white, draw=black, shape=circle, style=downground, scale=0.5]
\tikzstyle{state}=[fill=white, draw=black, style=triang, tikzit shape=rectangle]
\tikzstyle{state_small}=[fill=white, draw=black, style=triang_small, tikzit shape=rectangle]
\tikzstyle{state_lesspad}=[fill=white, draw=black, style=triang_lesssep, tikzit shape=rectangle]
\tikzstyle{kstate}=[fill=white, draw=black, style=kpoint, tikzit shape=rectangle]
\tikzstyle{kstateconj}=[fill=white, draw=black, style=kpoint conjugate, tikzit shape=rectangle]
\tikzstyle{kstateBIG}=[fill=white, draw=black, style=big kpoint, tikzit shape=rectangle]
\tikzstyle{effect}=[fill=white, draw=black, style=triangdag]
\tikzstyle{effect_small}=[fill=white, draw=black, style=triangdag_small]
\tikzstyle{keffect}=[fill=white, draw=black, style=kpoint adjoint]
\tikzstyle{keffectconj}=[fill=white, draw=black, style=kpoint transpose]
\tikzstyle{morphdag}=[style=mapdag]
\tikzstyle{morph}=[style=map]
\tikzstyle{morphtrans}=[style=maptrans]
\tikzstyle{morphconj}=[style=mapconj]
\tikzstyle{CPMmorph}=[style=dmap]
\tikzstyle{CPMmorphconj}=[style=dmapconj]
\tikzstyle{CPMmorphdag}=[style=dmapdag]
\tikzstyle{CPMmorphtrans}=[style=dmaptrans]
\tikzstyle{CPMstate}=[fill=white, draw=black, style=triang, doubled]
\tikzstyle{CPMstateBIG}=[fill=white, draw=black, style={triang_lesssep}, doubled]
\tikzstyle{CPMkstate}=[fill=white, draw=black, style=kpoint, tikzit shape=rectangle, doubled]
\tikzstyle{CPMkstateconj}=[fill=white, draw=black, style=kpoint conjugate, tikzit shape=rectangle, doubled]
\tikzstyle{CPMkstateBIG}=[fill=white, draw=black, style=big kpoint, tikzit shape=rectangle, doubled]
\tikzstyle{CPMkeffect}=[fill=white, draw=black, style=kpoint adjoint, doubled]
\tikzstyle{CPMkeffectconj}=[fill=white, draw=black, style=kpoint transpose, doubled]
\tikzstyle{UHfB}=[fill=white, draw=black, style=triangdag, doubled, inner sep=-2pt]
\tikzstyle{leak}=[style=tinypoint, regular polygon rotate=-90]
\tikzstyle{leakfill}=[style=tinypoint, regular polygon rotate=-90, fill=black]
\tikzstyle{Z}=[style=dot, fill=green]
\tikzstyle{X}=[style=dot, fill=red]
\tikzstyle{black_dot}=[style=dot, fill=black]
\tikzstyle{white_dot}=[style=dot, fill=white]
\tikzstyle{qblack_dot}=[style=ddot, fill=black]
\tikzstyle{qwhite_dot}=[style=ddot, fill=white]
\tikzstyle{whitephase}=[style=wphase dot, fill=white]
\tikzstyle{qredphase}=[style=phase dot, fill=red]
\tikzstyle{qgreenphase}=[style=phase dot, fill=green]
\tikzstyle{had}=[style=hadamard, doubled]
\tikzstyle{box}=[style=hadamard]
\tikzstyle{bigbox}=[style=hadamard, minimum height=6mm, minimum width=18mm]
\tikzstyle{WIDEBOI}=[style=hadamard, minimum height=6mm, minimum width=40mm]
\tikzstyle{WIDEstate}=[fill=white, draw=black, style=triang, tikzit shape=rectangle, minimum width=40mm]
\tikzstyle{bigboxCPM}=[style=hadamard, minimum height=6mm, minimum width=8mm, doubled]
\tikzstyle{antipode}=[style=anti]
\tikzstyle{hexspider}=[style=hex]
\tikzstyle{WIDEcutstate}=[style=state_doublecut, text depth = 2mm, minimum width = 8mm]
\tikzstyle{WIDERcutstate}=[style=state_doublecut, text depth = 0mm, minimum width = 36mm]
\tikzstyle{WIDEcutstateCPM}=[style=state_doublecut, text depth = 2mm, minimum width = 8mm, doubled]
\tikzstyle{WIDEcuteffect}=[style=effect_doublecut, text depth = 2mm, minimum width = 8mm]

\tikzstyle{dottededge}=[-, dotted]
\tikzstyle{double edge}=[-, style=doubled, draw=black, tikzit draw={rgb,255: red,191; green,0; blue,64}]

%% file: symbols/ZbwdotSym.tex
\begin{tikzpicture} [scale=1,transform shape] 

\def\deltax{0.3} 
\def\deltay{0.5} 


\node [dot, fill=\zcolour] (mult) at (0,0) {};

\end{tikzpicture}

%% file: symbols/ZbwcomultSym.tex
\begin{tikzpicture} [scale=0.6,transform shape] 

\def\deltax{0.3} 
\def\deltay{0.5} 


\node (mult_label_outl) at (-\deltax,+\deltay) {};
\node (mult_label_outr) at (+\deltax,+\deltay) {};
\node [dot, fill=\zcolour] (mult) at (0,0) {};
\node (mult_label_in) at (0,-\deltay) {};
\draw[-] [in=270,out=135] (mult) to (mult_label_outl);
\draw[-] [in=270,out=45] (mult) to (mult_label_outr);
\draw[-] (mult_label_in) to (mult);

\end{tikzpicture}

%% file: main_qpl.bbl
\begin{thebibliography}{10}
\providecommand{\bibitemdeclare}[2]{}
\providecommand{\surnamestart}{}
\providecommand{\surnameend}{}
\providecommand{\urlprefix}{Available at }
\providecommand{\url}[1]{\texttt{#1}}
\providecommand{\href}[2]{\texttt{#2}}
\providecommand{\urlalt}[2]{\href{#1}{#2}}
\providecommand{\doi}[1]{doi:\urlalt{http://dx.doi.org/#1}{#1}}
\providecommand{\bibinfo}[2]{#2}

\bibitemdeclare{inproceedings}{abramsky_coecke}
\bibitem{abramsky_coecke}
\bibinfo{author}{Samson \surnamestart Abramsky\surnameend} \&
  \bibinfo{author}{Bob \surnamestart Coecke\surnameend} (\bibinfo{year}{2004}):
  \emph{\bibinfo{title}{A categorical semantics of quantum protocols}}.
\newblock In: {\sl \bibinfo{booktitle}{Proceedings of the 19th Annual IEEE
  Symposium on Logic in Computer Science, 2004}}, pp.
  \bibinfo{pages}{415--425}, \doi{10.1109/LICS.2004.1319636}.

\bibitemdeclare{article}{ashoush}
\bibitem{ashoush}
\bibinfo{author}{Daniela \surnamestart Ashoush\surnameend} \&
  \bibinfo{author}{Bob \surnamestart Coecke\surnameend} (\bibinfo{year}{2016}):
  \emph{\bibinfo{title}{Dual Density Operators and Natural Language Meaning}}.
\newblock {\sl \bibinfo{journal}{EPTCS}} \bibinfo{volume}{221}, pp.
  \bibinfo{pages}{1--10}, \doi{10.4204/EPTCS.221.1}.

\bibitemdeclare{article}{beckmann_equivariant}
\bibitem{beckmann_equivariant}
\bibinfo{author}{Thorsten \surnamestart Beckmann\surnameend} \&
  \bibinfo{author}{Georg \surnamestart Oberdieck\surnameend}
  (\bibinfo{year}{2020}): \emph{\bibinfo{title}{Notes on equivariant
  categories}}.
\newblock {\sl \bibinfo{journal}{arXiv preprint}}
  \bibinfo{volume}{arXiv:2006.13626}.

\bibitemdeclare{book}{bourbaki_algebra}
\bibitem{bourbaki_algebra}
\bibinfo{author}{Nicolas \surnamestart Bourbaki\surnameend}
  (\bibinfo{year}{2003}): \emph{\bibinfo{title}{Algebra II}}.
\newblock \bibinfo{publisher}{Springer-Verlag Berlin Heidelberg},
  \doi{10.1007/978-3-642-61698-3}.

\bibitemdeclare{article}{coecke2016terminality}
\bibitem{coecke2016terminality}
\bibinfo{author}{Bob \surnamestart Coecke\surnameend} (\bibinfo{year}{2016}):
  \emph{\bibinfo{title}{Terminality implies no-signalling... and much more than
  that}}.
\newblock {\sl \bibinfo{journal}{New Generation Computing}}
  \bibinfo{volume}{34}(\bibinfo{number}{1-2}), pp. \bibinfo{pages}{69--85},
  \doi{10.1007/s00354-016-0201-6}.

\bibitemdeclare{article}{coecke_zx}
\bibitem{coecke_zx}
\bibinfo{author}{Bob \surnamestart Coecke\surnameend} \& \bibinfo{author}{Ross
  \surnamestart Duncan\surnameend} (\bibinfo{year}{2011}):
  \emph{\bibinfo{title}{Interacting quantum observables: categorical algebra
  and diagrammatics}}.
\newblock {\sl \bibinfo{journal}{New Journal of Physics}}
  \bibinfo{volume}{13}(\bibinfo{number}{043016}),
  \doi{10.1088/1367-2630/13/4/043016}.

\bibitemdeclare{article}{coecke_cp}
\bibitem{coecke_cp}
\bibinfo{author}{Bob \surnamestart Coecke\surnameend}, \bibinfo{author}{Chris
  \surnamestart Heunen\surnameend} \& \bibinfo{author}{Aleks \surnamestart
  Kissinger\surnameend} (\bibinfo{year}{2016}):
  \emph{\bibinfo{title}{Categories of quantum and classical channels}}.
\newblock {\sl \bibinfo{journal}{Quantum Inf. Process.}} \bibinfo{volume}{15},
  p. \bibinfo{pages}{5179–5209}, \doi{10.1007/s11128-014-0837-4}.

\bibitemdeclare{book}{coecke_kissinger_2017}
\bibitem{coecke_kissinger_2017}
\bibinfo{author}{Bob \surnamestart Coecke\surnameend} \& \bibinfo{author}{Aleks
  \surnamestart Kissinger\surnameend} (\bibinfo{year}{2017}):
  \emph{\bibinfo{title}{Picturing Quantum Processes: A First Course in Quantum
  Theory and Diagrammatic Reasoning}}.
\newblock \bibinfo{publisher}{Cambridge University Press},
  \doi{10.1017/9781316219317}.

\bibitemdeclare{article}{coecke_double}
\bibitem{coecke_double}
\bibinfo{author}{Bob \surnamestart Coecke\surnameend} \&
  \bibinfo{author}{Konstantinos \surnamestart Meichanetzidis\surnameend}
  (\bibinfo{year}{2020}): \emph{\bibinfo{title}{Meaning Updating of Density
  Matrices}}.
\newblock {\sl \bibinfo{journal}{arXiv preprint}}
  \bibinfo{volume}{arXiv:2001.00862v1}.

\bibitemdeclare{article}{coecke_bases}
\bibitem{coecke_bases}
\bibinfo{author}{Bob \surnamestart Coecke\surnameend}, \bibinfo{author}{Dusko
  \surnamestart Pavlovic\surnameend} \& \bibinfo{author}{Jamie \surnamestart
  Vicary\surnameend} (\bibinfo{year}{2013}): \emph{\bibinfo{title}{A new
  description of orthogonal bases}}.
\newblock {\sl \bibinfo{journal}{Mathematical Structures in Computer Science}}
  \bibinfo{volume}{23}(\bibinfo{number}{3}), pp. \bibinfo{pages}{555--567},
  \doi{10.1017/S0960129512000047}.

\bibitemdeclare{article}{coecke_environment}
\bibitem{coecke_environment}
\bibinfo{author}{Bob \surnamestart Coecke\surnameend} \& \bibinfo{author}{Simon
  \surnamestart Perdrix\surnameend} (\bibinfo{year}{2012}):
  \emph{\bibinfo{title}{Environment and classical channels in categorical
  quantum mechanics}}.
\newblock {\sl \bibinfo{journal}{Logical Methods in Computer Science}}
  \bibinfo{volume}{8}, \doi{10.2168/LMCS-8(4:14)2012}.

\bibitemdeclare{article}{coecke_classicality}
\bibitem{coecke_classicality}
\bibinfo{author}{Bob \surnamestart Coecke\surnameend}, \bibinfo{author}{John
  \surnamestart Selby\surnameend} \& \bibinfo{author}{Sean \surnamestart
  Tull\surnameend} (\bibinfo{year}{2018}): \emph{\bibinfo{title}{Two Roads to
  Classicality}}.
\newblock {\sl \bibinfo{journal}{EPTCS}} \bibinfo{volume}{266}, p.
  \bibinfo{pages}{104–118}, \doi{10.4204/eptcs.266.7}.

\bibitemdeclare{article}{Cunningham_cp}
\bibitem{Cunningham_cp}
\bibinfo{author}{Oscar \surnamestart Cunningham\surnameend} \&
  \bibinfo{author}{Chris \surnamestart Heunen\surnameend}
  (\bibinfo{year}{2015}): \emph{\bibinfo{title}{Axiomatizing complete
  positivity}}.
\newblock {\sl \bibinfo{journal}{EPTCS}} \bibinfo{volume}{195}, p.
  \bibinfo{pages}{148–157}, \doi{10.4204/eptcs.195.11}.

\bibitemdeclare{article}{elagin_equivariant}
\bibitem{elagin_equivariant}
\bibinfo{author}{Alexey \surnamestart Elagin\surnameend}
  (\bibinfo{year}{2015}): \emph{\bibinfo{title}{On equivariant triangulated
  categories}}.
\newblock {\sl \bibinfo{journal}{arXiv preprint}}
  \bibinfo{volume}{arXiv:1403.7027}.

\bibitemdeclare{article}{ellison1}
\bibitem{ellison1}
\bibinfo{author}{William \surnamestart Ellison\surnameend}
  (\bibinfo{year}{1970-1971}): \emph{\bibinfo{title}{Waring's Problem for
  Fields}}.
\newblock {\sl \bibinfo{journal}{S\'eminaire de th\'eorie des nombres de
  Bordeaux}}, pp. \bibinfo{pages}{1--8}.
\newblock \urlprefix\url{www.numdam.org/item/STNB_1970-1971____A8_0/}.

\bibitemdeclare{article}{ellison2}
\bibitem{ellison2}
\bibinfo{author}{William \surnamestart Ellison\surnameend}
  (\bibinfo{year}{2013}): \emph{\bibinfo{title}{Waring's problem for fields}}.
\newblock {\sl \bibinfo{journal}{arXiv preprint}}
  \bibinfo{volume}{arXiv:1303.4818}.

\bibitemdeclare{article}{ganter_symmetric}
\bibitem{ganter_symmetric}
\bibinfo{author}{Nora \surnamestart Ganter\surnameend} \&
  \bibinfo{author}{Mikhail \surnamestart Kapranov\surnameend}
  (\bibinfo{year}{2014}): \emph{\bibinfo{title}{Symmetric and Exterior Powers
  of Categories}}.
\newblock {\sl \bibinfo{journal}{Transformation Groups}} \bibinfo{volume}{19},
  pp. \bibinfo{pages}{57--103}, \doi{10.1007/s00031-014-9255-z}.

\bibitemdeclare{article}{gogioso_fantastic}
\bibitem{gogioso_fantastic}
\bibinfo{author}{Stefano \surnamestart Gogioso\surnameend}
  (\bibinfo{year}{2017}): \emph{\bibinfo{title}{Fantastic Quantum Theories and
  Where to Find Them}}.
\newblock {\sl \bibinfo{journal}{arXiv Preprint}}
  \bibinfo{volume}{arXiv:1703.10576}.

\bibitemdeclare{article}{gogioso_cpm}
\bibitem{gogioso_cpm}
\bibinfo{author}{Stefano \surnamestart Gogioso\surnameend}
  (\bibinfo{year}{2019}): \emph{\bibinfo{title}{Higher-order CPM
  Constructions}}.
\newblock {\sl \bibinfo{journal}{EPTCS}} \bibinfo{volume}{287}, pp.
  \bibinfo{pages}{145--162}, \doi{10.4204/EPTCS.287.8}.

\bibitemdeclare{article}{gogioso_hiddensubgroup}
\bibitem{gogioso_hiddensubgroup}
\bibinfo{author}{Stefano \surnamestart Gogioso\surnameend} \&
  \bibinfo{author}{Aleks \surnamestart Kissinger\surnameend}
  (\bibinfo{year}{2017}): \emph{\bibinfo{title}{Fully graphical treatment of
  the quantum algorithm for the Hidden Subgroup Problem}}.
\newblock {\sl \bibinfo{journal}{arXiv preprint}}
  \bibinfo{volume}{arXiv:1701.08669}.

\bibitemdeclare{inproceedings}{gogioso_hyper}
\bibitem{gogioso_hyper}
\bibinfo{author}{Stefano \surnamestart Gogioso\surnameend} \&
  \bibinfo{author}{Carlo~Maria \surnamestart Scandolo\surnameend}
  (\bibinfo{year}{2019}): \emph{\bibinfo{title}{Density Hypercubes, Higher
  Order Interference and Hyper-decoherence: A Categorical Approach}}.
\newblock In: {\sl \bibinfo{booktitle}{Quantum Interaction}},
  \bibinfo{publisher}{Springer International Publishing}, pp.
  \bibinfo{pages}{141--160}, \doi{10.1007/978-3-030-35895-2\textunderscore 10}.

\bibitemdeclare{article}{gogioso2015fourier}
\bibitem{gogioso2015fourier}
\bibinfo{author}{Stefano \surnamestart Gogioso\surnameend} \&
  \bibinfo{author}{William \surnamestart Zeng\surnameend}
  (\bibinfo{year}{2015}): \emph{\bibinfo{title}{Fourier transforms from
  strongly complementary observables}}.
\newblock {\sl \bibinfo{journal}{arXiv Preprint}}
  \bibinfo{volume}{arXiv:1501.04995}.

\bibitemdeclare{article}{hefford_hypercubes}
\bibitem{hefford_hypercubes}
\bibinfo{author}{James \surnamestart Hefford\surnameend} \&
  \bibinfo{author}{Stefano \surnamestart Gogioso\surnameend}
  (\bibinfo{year}{2020}): \emph{\bibinfo{title}{Hyper-decoherence in Density
  Hypercubes}}.
\newblock {\sl \bibinfo{journal}{arXiv preprint}}
  \bibinfo{volume}{arXiv:2003.08318}.

\bibitemdeclare{book}{heunen_categories}
\bibitem{heunen_categories}
\bibinfo{author}{Chris \surnamestart Heunen\surnameend} \&
  \bibinfo{author}{Jamie \surnamestart Vicary\surnameend}
  (\bibinfo{year}{2019}): \emph{\bibinfo{title}{Categories for Quantum Theory:
  An Introduction}}.
\newblock \bibinfo{publisher}{Oxford University Press},
  \doi{10.1093/oso/9780198739623.001.0001}.

\bibitemdeclare{book}{jacobson_algebra}
\bibitem{jacobson_algebra}
\bibinfo{author}{Nathan \surnamestart Jacobson\surnameend}
  (\bibinfo{year}{1989}): \emph{\bibinfo{title}{Basic Algebra II: Second
  Edition}}.
\newblock \bibinfo{publisher}{W. H. Freeman and Company}.

\bibitemdeclare{article}{lack_props}
\bibitem{lack_props}
\bibinfo{author}{Stephen \surnamestart Lack\surnameend} (\bibinfo{year}{2004}):
  \emph{\bibinfo{title}{Composing PROPs}}.
\newblock {\sl \bibinfo{journal}{Theory and Applications of Categories}}
  \bibinfo{volume}{13}(\bibinfo{number}{9}), pp. \bibinfo{pages}{147--163}.

\bibitemdeclare{article}{lee_interference}
\bibitem{lee_interference}
\bibinfo{author}{Ciar\'an~M. \surnamestart Lee\surnameend} \&
  \bibinfo{author}{John~H. \surnamestart Selby\surnameend}
  (\bibinfo{year}{2017}): \emph{\bibinfo{title}{Higher-Order Interference in
  Extensions of Quantum Theory}}.
\newblock {\sl \bibinfo{journal}{Found Phys}} \bibinfo{volume}{47}, pp.
  \bibinfo{pages}{89--112}, \doi{10.1007/s10701-016-0045-4}.

\bibitemdeclare{article}{lee_nogo}
\bibitem{lee_nogo}
\bibinfo{author}{Ciar\'an~M. \surnamestart Lee\surnameend} \&
  \bibinfo{author}{John~H. \surnamestart Selby\surnameend}
  (\bibinfo{year}{2018}): \emph{\bibinfo{title}{A no-go theorem for theories
  that decohere to quantum mechanics}}.
\newblock {\sl \bibinfo{journal}{Proc. R. Soc. A}}
  \bibinfo{volume}{474}(\bibinfo{number}{20170732}),
  \doi{10.1098/rspa.2017.0732}.

\bibitemdeclare{article}{piedeleu}
\bibitem{piedeleu}
\bibinfo{author}{Robin \surnamestart Piedeleu\surnameend},
  \bibinfo{author}{Dimitri \surnamestart Kartsaklis\surnameend},
  \bibinfo{author}{Bob \surnamestart Coecke\surnameend} \&
  \bibinfo{author}{Mehrnoosh \surnamestart Sadrzadeh\surnameend}
  (\bibinfo{year}{2015}): \emph{\bibinfo{title}{Open System Categorical Quantum
  Semantics in Natural Language Processing}}.
\newblock {\sl \bibinfo{journal}{arXiv preprint}}
  \bibinfo{volume}{arXiv:1502.00831}.

\bibitemdeclare{book}{rotman_algebra}
\bibitem{rotman_algebra}
\bibinfo{author}{Joseph~J. \surnamestart Rotman\surnameend}
  (\bibinfo{year}{2002}): \emph{\bibinfo{title}{Advanced Modern Algebra}}.
\newblock \bibinfo{publisher}{Prentice Hall}.

\bibitemdeclare{article}{Selby_coherence}
\bibitem{Selby_coherence}
\bibinfo{author}{John~H. \surnamestart Selby\surnameend} \&
  \bibinfo{author}{Ciar{\'{a}}n~M. \surnamestart Lee\surnameend}
  (\bibinfo{year}{2020}): \emph{\bibinfo{title}{Compositional resource theories
  of coherence}}.
\newblock {\sl \bibinfo{journal}{{Quantum}}} \bibinfo{volume}{4}, p.
  \bibinfo{pages}{319}, \doi{10.22331/q-2020-09-11-319}.

\bibitemdeclare{article}{selinger_cpm}
\bibitem{selinger_cpm}
\bibinfo{author}{Peter \surnamestart Selinger\surnameend}
  (\bibinfo{year}{2007}): \emph{\bibinfo{title}{Dagger Compact Closed
  Categories and Completely Positive Maps}}.
\newblock {\sl \bibinfo{journal}{Electronic Notes in Theoretical Computer
  Science}} \bibinfo{volume}{170}, pp. \bibinfo{pages}{139--163},
  \doi{10.1016/j.entcs.2006.12.018}.

\bibitemdeclare{article}{selinger_idempotents}
\bibitem{selinger_idempotents}
\bibinfo{author}{Peter \surnamestart Selinger\surnameend}
  (\bibinfo{year}{2008}): \emph{\bibinfo{title}{Idempotents in Dagger
  Categories: (Extended Abstract)}}.
\newblock {\sl \bibinfo{journal}{Electronic Notes in Theoretical Computer
  Science}} \bibinfo{volume}{210}, pp. \bibinfo{pages}{107--122},
  \doi{10.1016/j.entcs.2008.04.021}.

\bibitemdeclare{article}{shinder_groupactions}
\bibitem{shinder_groupactions}
\bibinfo{author}{Evgeny \surnamestart Shinder\surnameend}
  (\bibinfo{year}{2018}): \emph{\bibinfo{title}{Group actions on categories and
  Elagin’s theorem revisited}}.
\newblock {\sl \bibinfo{journal}{European Journal of Mathematics}}
  \bibinfo{volume}{4}, pp. \bibinfo{pages}{413--422},
  \doi{10.1007/s40879-017-0150-8}.

\bibitemdeclare{article}{siegel}
\bibitem{siegel}
\bibinfo{author}{Carl \surnamestart Siegel\surnameend} (\bibinfo{year}{1921}):
  \emph{\bibinfo{title}{Darstellung total positiver Zahlen durch Quadrate}}.
\newblock {\sl \bibinfo{journal}{Math Z}} \bibinfo{volume}{11}, pp.
  \bibinfo{pages}{246--275}, \doi{10.1007/BF01203627}.

\bibitemdeclare{inproceedings}{vicary_topology}
\bibitem{vicary_topology}
\bibinfo{author}{Jamie \surnamestart Vicary\surnameend} (\bibinfo{year}{2013}):
  \emph{\bibinfo{title}{Topological Structure of Quantum Algorithms}}.
\newblock In: {\sl \bibinfo{booktitle}{2013 28th Annual ACM/IEEE Symposium on
  Logic in Computer Science}}, pp. \bibinfo{pages}{93--102},
  \doi{10.1109/LICS.2013.14}.

\bibitemdeclare{article}{coecke_dilation}
\bibitem{coecke_dilation}
\bibinfo{author}{Maaike \surnamestart Zwart\surnameend} \& \bibinfo{author}{Bob
  \surnamestart Coecke\surnameend} (\bibinfo{year}{2018}):
  \emph{\bibinfo{title}{Double Dilation $\neq$ Double Mixing}}.
\newblock {\sl \bibinfo{journal}{EPTCS}} \bibinfo{volume}{266}, pp.
  \bibinfo{pages}{133--146}, \doi{10.4204/EPTCS.266.9}.

\bibitemdeclare{article}{zyczkowski_quartic}
\bibitem{zyczkowski_quartic}
\bibinfo{author}{Karol \surnamestart \.{Z}yczkowski\surnameend}
  (\bibinfo{year}{2008}): \emph{\bibinfo{title}{Quartic quantum theory: an
  extension of the standard quantum mechanics}}.
\newblock {\sl \bibinfo{journal}{J. Phys. A: Math. Theor.}}
  \bibinfo{volume}{41}(\bibinfo{number}{355302}),
  \doi{10.1088/1751-8113/41/35/355302}.

\end{thebibliography}
